\documentclass[aps,prmaterials,twocolumn,longbibliography,nobibnotes,showkeys]{revtex4-2}

\usepackage{graphicx}
\usepackage{dcolumn}
\usepackage{bm}
\usepackage{booktabs}
\usepackage{amssymb}
\usepackage{amsmath}

\begin{document}
\title{First-principles identification of optically efficient erbium centers in GaAs}
\author{Khang Hoang}
\email{khang.hoang@ndsu.edu}
\affiliation{Center for Computationally Assisted Science and Technology \& Department of Physics, North Dakota State University, Fargo, North Dakota 58108, United States}

\date{\today}

\begin{abstract}

Gallium arsenide (GaAs) doped with erbium (Er), a material of interest for optoelectronics and quantum information, has been studied for decades. Yet the formation of Er luminescence centers in the semiconductor host and their properties are still not well understood. Here we present a systematic investigation of Er-related defects in GaAs, including defect complexes consisting of Er and native point defects or oxygen impurities, using first-principles hybrid-functional defect calculations. We find that these defects have electronic structure and energetics that are generally asymmetric with respect to n- and p-type doping and tend to favor electron trapping. On the basis of the calculated defect levels, formation energies, and nonradiative carrier capture coefficients, we identify Er-related defects that are efficient as trap-assisted nonradiative recombination centers for Er$^{3+}$ excitation under host photoexcitation or via minority carrier injection. Our results provide an understanding for why a particular defect center with Er coupled to two oxygen atoms is most efficient, and for the effects of n- and p-type doping and of the Er/O ratio on the formation of optically active Er centers and on the Er luminescence observed in experiments.  

\end{abstract}


\maketitle


\section{Introduction}\label{sec;intro}

Semiconductors doped with rare-earth (RE) ions have long been of interest for optoelectronics and spintronics~\cite{ODonnell2010Book}. More recently, they have also been considered for quantum information applications, e.g., as single-photon emitters and spin-photon interfaces~\cite{Yin2013Nature,Ogawa2020APL,Higashi2020JAP,Fang2023OC,Fang2025JJAP}. The materials exhibit features similar to those found in RE-doped complex oxide insulators that have been used in quantum computing and quantum memory experiments~\cite{Thiel2011JL,Kunkel2018ZAAC}. These include sharp and stable intra-$4f$-shell optical transitions and excellent coherence properties, resulting from the RE $4f$-electron core being well shielded by the outer $5s^2$ and $5p^6$ electron shells. Yet the advantage of using semiconductor hosts is that the materials can also be controlled electrically, as opposed to just optically. 

Among the RE-doped semiconductors, GaAs doped with trivalent erbium (Er$^{3+}$ $4f^{11}$) has been studied extensively experimentally~\cite{Ennen1983APL,Pomrenke1986JAP,Smith1987APL,Ennen1987JAP,Thonke1988JPC,Takahei1989JAP,Kozanecki1991SSC,Nakata1992APL,Nakata1995JAP,Alves1998NIMPR,Wahl1999NIMPR}, especially when codoped with oxygen~\cite{Nakagome1988APL,Taguchi1993MRS,Takahei1993JAP,Takahei1994JJAP,Takahei1994JAP,Kaczanowski1996NIM,Ofuchi2000ME,Takahei1997JAP,Takahei1995JAP,Takahei1995JAP77,Ishiyama1998JAP,Haase1996MRS,Hogg1996JAP,Hogg1996APL,Culp1998JAP,Cederberg1999JAP,Fujiwara1999PB,Koizumi2003PB,Koizumi2003APL,Fujiwara2005MT,Taguchi1996JAP,Takahei1996MRS,Hogg1997PRB,Nakamura2006PB,Ishii2014JAP,Katsuno2011JL,Elmasry2014JAP,Maltez2004NIMPR}. The Er$^{3+}$ ion offers luminescence peaks at the telecommunication wavelength (around 1.54 $\mu$m) originating from the $^4I_{13/2}$ $\longrightarrow$ $^4I_{15/2}$ transition within the $4f$ multiplet. Fujiwara and collaborators successfully fabricated light-emitting diodes based on (Er,O)-doped GaAs operating at room temperature~\cite{Koizumi2003PB,Koizumi2003APL,Fujiwara2005MT}. They were also able to couple the most optically efficient (Er,O)-related defect center (referred to as ``Er-2O'') to photonic crystal nanocavities or microdisk resonators and achieved up to about tenfold enhancement of Er luminescence via the Purcell effect~\cite{Ogawa2020APL,Higashi2020JAP,Fang2023OC,Fang2025JJAP}. Such a progress constitutes a major step toward realizing single-photon emitters or laser devices based on the material. As further discussed below, however, the formation of Er-related luminescence centers in GaAs and their properties are still not well understood. In general, whether an RE dopant is being harnessed for traditional optoelectronic applications or for novel quantum technologies, having a detailed understanding of the interaction between the RE ion and the semiconductor host (including its native point defects, unintentional impurities, and codopants) and of the RE luminescence will assist in materials design and thus facilitate further advances in these areas. 

\begin{figure}
\vspace{0.2cm}
\includegraphics*[width=0.95\linewidth]{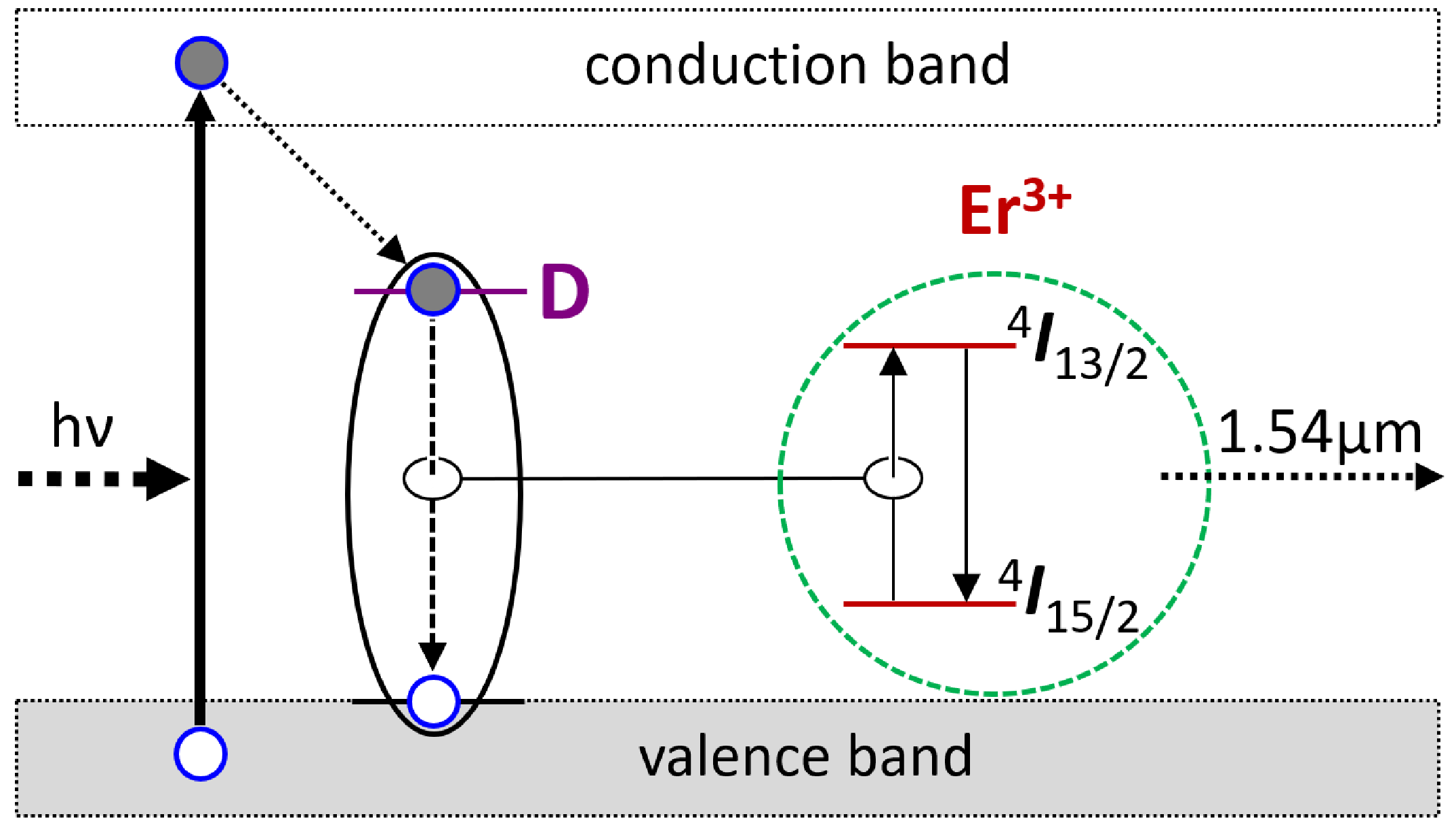}
\caption{Schematic illustration of indirect Er$^{3+}$ excitation. Following band-to-band excitation of the semiconductor host, an electron is excited from the valence band to the conduction band. The excited electron is then trapped at a defect center D before recombining nonradiatively with a hole and the recombination energy is transferred into the Er$^{3+}$ $4f$-electron core. The same mechanism, {\it mutatis mutandis}, is applied to host photoexcitation involving hole trapping or when host photoexcitation is replaced with minority carrier injection.}
\label{fig;excitation}
\end{figure} 

A RE luminescence center can be optically excited by direct (resonant) intra-$4f$-shell excitation or by indirect (nonresonant) excitation of the semiconductor host. In the former the excitation energy is directly absorbed into the $4f$-electron core (which is an inefficient process due to the parity-forbidden nature of the intra-$4f$-shell optical transitions), whereas in the latter it is indirectly transferred from the host. The host excitation mechanism is expected to be mediated by defects which act as trap-assisted nonradiative recombination centers; see Fig.~\ref{fig;excitation}. An electron trapped at a defect center D, for instance, can recombine nonradiatively with a hole from the valence band or from some acceptor level. The recombination energy is transferred into the $4f$-electron core which excites an electron from one $4f$ level (e.g., $^4I_{15/2}$ as seen in Fig.~\ref{fig;excitation}) to another ($^4I_{13/2}$ or higher) via an Auger-type process. The excited electron in the case of Er$^{3+}$ can then relax from $^4I_{13/2}$ (excited state) to $^4I_{15/2}$ (ground state) and emit a photon at about 1.54 $\mu$m. The energy mismatch between the nonradiative recombination at D and the intra-$4f$-excitation may be assumed to be compensated by multiphonon processes~\cite{Taguchi1996JAP}. Instead of excited electrons from photoexcitation of the host, electrons can be provided via minority carrier injection in a p-n junction. The defect center D, if efficient in host photoexcitation, should also be efficient when carriers are injected. Here we focus on D as an electron trap, but a similar description can be made regarding hole trapping.
 
For the energy transfer from the host to the RE $4f$-electron core to be efficient, the recombination center D needs to be in close proximity of the RE ion and the energy mismatch to be small~\cite{Takahei1996MRS,Gregorkiewicz1999MRSBulletin}. For that reason, efforts have been focused on searching for recombination centers D in which the RE ion is an integral part of D, i.e., D is a RE-related single defect or defect complex (hereafter commonly referred to as RE-related defect)~\cite{Hoang2015RRL,Hoang2016RRL,Hoang201JAP,Hoang2021PRM,Hoang2022PRM}. In such a defect, the constituent(s) other than the RE ion include localized carrier(s), native point defect(s), unintentional impuriti(es), and/or codopant(s). Strong defect--defect interaction within the RE-related defect is utilized to enhance energy transfer efficiency. Requirements for such a defect are that (i) the Er-related defect induces charge-state defect transition level(s) in the band gap of the host material and (ii) the transition level(s) are suitable for high-energy recombination (e.g., in the case of Er$^{3+}$ excitation, the nonradiative recombination energy should be at least about 0.81 eV, which is the $^4I_{15/2}$ $\longrightarrow$ $^4I_{13/2}$ excitation energy), but not much higher than the intra-$4f$-excitation energy to keep the energy mismatch small. Other features include that (iii) the carrier-capturing defect configuration is non-repulsive to minority carriers and (iv) the defect has a relatively low formation energy. Condition (iv) is to ensure that the defect can be formed and stable, and that it can occur with a sizable concentration (although this last bit is not strictly required as only a single defect may be needed, e.g., when it is coupled to photonic crystal nanocavities for luminescence enhancement). These criteria will be made clearer in our discussions in Sec.~\ref{sec;nonrad}.

Experimentally, Er-doped GaAs samples are often obtained by ion implantation, metal-organic chemical vapor deposition (MOCVD), or molecular beam epitaxy (MBE). Ennen et al.~\cite{Ennen1983APL} were the first who observed fine-structured luminescence bands with sharp lines in the 1.54-$\mu$m region in Er-implanted GaAs even at room temperature. There have been multiple experimental reports of Er at the interstitial sites~\cite{Kozanecki1991SSC,Nakata1992APL,Nakata1995JAP,Alves1998NIMPR,Wahl1999NIMPR}. Kozanecki et al.~\cite{Kozanecki1991SSC}, for example, found that as-implanted Er atoms first occupy the interstitial sites and then move to the substitutional sites after thermal annealing, and the optical activity of the Er ions (under host photoexcitation) disappears entirely after they have been located on substitutional sites. The interstitial Er defects are thus metastable. Notably, Er-doped GaAs samples often contain oxygen either as an unintentional impurity or codopant and, in the presence of oxygen, Er was found to be mainly (approximately) at the substitutional lattice site~\cite{Takahei1994JAP,Kaczanowski1996NIM,Ofuchi2000ME,Takahei1997JAP}. Codoping with oxygen also leads to sharper, simpler, and intense Er photoluminescence (PL) spectra~\cite{Nakagome1988APL,Taguchi1993MRS,Takahei1993JAP,Takahei1994JJAP,Takahei1994JAP,Kaczanowski1996NIM,Ofuchi2000ME,Takahei1997JAP,Takahei1995JAP,Takahei1995JAP77,Ishiyama1998JAP,Haase1996MRS,Hogg1996JAP,Takahei1996MRS,Hogg1997PRB,Hogg1996APL,Culp1998JAP,Cederberg1999JAP,Fujiwara1999PB,Koizumi2003PB,Koizumi2003APL,Fujiwara2005MT,Nakamura2006PB,Katsuno2011JL,Ishii2014JAP,Elmasry2014JAP,Maltez2004NIMPR}.

Multiple Er luminescence centers have been observed in Er-doped GaAs (either with deliberate oxygen codoping or not)~\cite{Pomrenke1986JAP,Smith1987APL,Ennen1987JAP,Takahei1995JAP,Takahei1995JAP77,Takahei1997JAP,Ishiyama1998JAP}. Takahei and Taguchi~\cite{Takahei1995JAP}, for example, observed three types of Er centers in (Er,O)-doped GaAs obtained by MOCVD: (i) Type I, identified as an Er atom coupled to two oxygen atoms (i.e., Er-2O, with the $C_{2v}$ symmetry) first by Takahei et al.~\cite{Takahei1994JAP}, shows sharp and simple PL spectra with a high intensity under host photoexcitation. (ii) Type II shows sharp and simple spectra but only under direct intra‐$4f$‐shell photoexcitation and not under host photoexcitation. Several such centers were found at sizable concentrations, probably higher than that of the Er-2O center. (iii) Type III, suggested to be related to ``aggregates of Er complexes,'' shows complex spectra under direct intra‐$4f$‐shell photoexcitation with a specific photon energy. This type of Er centers also shows luminescence under host photoexcitation but with a much lower intensity than Er-2O. Overall, Er-2O was found to dominate the Er luminescence spectrum under host photoexcitation~\cite{Takahei1997JAP}, although it may not be the predominant Er center. Deviation from the Er:2O ratio, e.g., having excessive oxygen, has been shown to lead to performance degradation~\cite{Maltez2004NIMPR}.  

It has generally been suggested that Er-2O acts as an electron trap~\cite{Takahei1996MRS,Fujiwara2005MT,Nakamura2006PB,Ishii2014JAP}, although the energy level of the trap is not entirely clear. Hogg et al.~\cite{Hogg1996JAP} argued that the trap depth should be in the 0.82--1.22 eV region [above the valence-band maximum (VBM), if electron trapping is assumed as it was not possible to determine if it was an electron or hole trap]. The numbers were deduced from the fact that (i) the depth trap must be higher than the $^4I_{15/2}$$\longrightarrow$$^4I_{13/2}$ excitation energy and (ii) no evidence for an Er-2O related trap level was found by PL excitation spectroscopy in the energy region from 1.22 eV to the band gap~\cite{Hogg1996JAP}. On the other hand and based on an analysis of the PL thermal quenching, an Er-2O related trap level was estimated to be at 0.42 eV~\cite{Taguchi1996JAP,Takahei1996MRS} or 0.375$\pm$0.05 eV~\cite{Hogg1997PRB} [below the conduction-band minimum (CBM), if electron trapping is assumed]. Effects of n- or p-type doping (e.g., via S doping at the As site or Zn doping at the Ga site, respectively) has also been investigated and it was found that n-type doping greatly suppressed the formation of Er-2O~\cite{Fujiwara2005MT} and the Er PL~\cite{Culp1998JAP,Cederberg1999JAP}.

Despite the vast experimental literature, computational investigations into Er-related defects in GaAs have been limited. Taguchi and Ohno~\cite{Taguchi1997PRB} and Svane et al.~\cite{Svane2006} reported defect levels associated with isolated substitutional and interstitial Er defects. Coutinho et al.~\cite{Coutinho2004APL} reported results for several important (Er,O)-related defect complexes. All these studies, however, were based on density-functional theory (DFT) within the local-density approximation (LDA)~\cite{Taguchi1997PRB,Coutinho2004APL} or self-interaction corrected LDA~\cite{Svane2006} (and often with small supercell models~\cite{Taguchi1997PRB,Svane2006}) which are known to have limited predictive power in the study of defects in semiconductors and insulators~\cite{Freysoldt2014RMP,Hoang2021PRM,Hoang2026JPCM}. Yet the above discussion and literature review leave us with many questions: How do Er-2O and other Er-related defects form in GaAs and what are their atomic and electronic structure, including trap depths induced by these defects? What make some of these defects optically efficient? What is exactly Er-2O and why is it the best Er luminescence center under host photoexcitation or via minority carrier injection? There has also been a complete lack of understanding of the effects of n- and p-type doping and of the Er/O ratio on the formation of optically active Er centers and on the Er PL. 

We herein report a systematic investigation of Er-related defects in GaAs, including defect complexes consisting of Er and native point defects or oxygen impurities, using state-of-the-art first-principles defect calculations~\cite{Freysoldt2014RMP}. Defect centers suitable for Er$^{3+}$ excitation, via either photoexcitation of the semiconductor host or minority carrier injection, are identified on the basis of the criteria we discussed earlier, and first-principles calculations of nonradiative carrier capture coefficients~\cite{Alkauskas2014PRB,Turiansky2021CPC} are then carried out to assess their efficiency as trap-assisted recombination centers. With regard to the excitation mechanism illustrated in Fig.~\ref{fig;excitation} and discussed above, we thus focus primarily on carrier capture at a defect center D and address neither the energy transfer process nor the intra-$4f$-shell transitions directly.

\section{Methodology}\label{sec;method} 

First-principles total-energy electronic structure calculations are based on DFT with the Heyd-Scuseria-Ernzerhof (HSE) functional~\cite{heyd:8207}, the projector augmented wave (PAW) method~\cite{PAW2}, and a plane-wave basis set, as implemented in \textsc{vasp}~\cite{VASP2}. We use the standard PAW potentials in the \textsc{vasp} database which treat Ga $4s^24p^1$, As $4s^24p^3$, Er $5s^25p^64f^{12}6s^2$, and O $2s^22p^4$ explicitly as valence electrons and the rest as core electrons. The Hartree-Fock mixing parameter is set to 0.28 and the screening length to the default value of 10 {\AA}; the plane-wave basis-set cutoff is set to 400 eV; and spin polarization is included. This choice of the hybrid density functional and associated parameters for zincblende GaAs results in a band gap of 1.51 eV, a lattice constant of 5.65 {\AA}, and a total static dielectric constant of 12.19~\cite{Hoang2026JPCM}, all at 0 K. Defects in the GaAs host are simulated using 3$\times$3$\times$3 (216-atom) cubic supercells. Integrations over the Brillouin zone are carried out using the $\Gamma$ point (see below for more on the choice of \textbf{k}-points). All structural relaxations are performed with HSE and the force threshold is chosen to be 0.01 eV/{\AA}. More computational details can be found in Ref.~\citenum{Hoang2026JPCM}. 

The formation energy of a defect X with charge state $q$ (with respect to the host lattice) is defined as~\cite{walle:3851,Freysoldt2014RMP}     
\begin{align}\label{eq:eform}
E^f({\mathrm{X}}^q)&=&E_{\mathrm{tot}}({\mathrm{X}}^q)-E_{\mathrm{tot}}({\mathrm{bulk}}) -\sum_{i}{n_i\mu_i} \\ %
\nonumber &&+~q(E_{\mathrm{v}}+\mu_{e})+ \Delta^q ,
\end{align}
where $E_{\mathrm{tot}}(\mathrm{X}^{q})$ and $E_{\mathrm{tot}}(\mathrm{bulk})$ are the total energies of the defect-containing and bulk supercells; $n_{i}$ is the number of atoms of species $i$ that have been added ($n_{i}>0$) or removed ($n_{i}<0$) to form the defect; $\mu_{i}$ is the atomic chemical potential, representing the energy of the reservoir with which atoms are being exchanged. $\mu_{e}$ is the electronic chemical potential, i.e., the Fermi level, representing the energy of the electron reservoir, referenced to the VBM of the host ($E_{\mathrm{v}}$). Finally, $\Delta^q$ is the correction term to align the electrostatic potentials of the bulk and defect-containing supercells and to account for finite-size effects on the total energies of charged defects~\cite{Freysoldt,Freysoldt11}.

The chemical potentials of Ga and As are referenced to the total energy per atom of bulk Ga and As and vary over a range determined by the formation enthalpy of GaAs: $\mu_{\rm Ga} + \mu_{\rm As} = \Delta H ({\rm GaAs}) = -0.96$ eV (per formula unit and at 0 K)~\cite{Hoang2026JPCM}. In the following presentation, we focus on the extreme Ga-rich and As-rich conditions which correspond to $\mu_{\rm Ga} = 0$ eV and $\mu_{\rm As} = 0$ eV, respectively. The chemical potential of oxygen ($\mu_{\rm O}$) is assumed to be limited by the formation of $\beta$-Ga$_2$O$_3$ ($\Delta H = -10.15$ eV)~\cite{Hoang2026JPCM}, and of erbium ($\mu_{\rm Er}$) by the formation of ErAs ($\Delta H = -3.58$ eV at 0 K, compared to $-3.28 \pm 0.14$ eV measured at 405--445$^\circ$ C~\cite{Hanks1967TFS}). ErAs sometimes occurs as an aggregate in Er-implanted GaAs~\cite{Alves1998NIMPR}.  

The thermodynamic transition level between charge states $q_1$ and $q_2$ of a defect, $\epsilon(q_1/q_2)$, is defined as~\cite{Freysoldt2014RMP}
\begin{equation}\label{eq;tl}
\epsilon(q_1/q_2) = \frac{E^f(X^{q_1}; \mu_e=0)-E^f(X^{q_2}; \mu_e=0)}{q_2 - q_1},
\end{equation}
where $E^f(X^{q}; \mu_e=0)$ is the formation energy of the defect X in charge state $q$ when the Fermi level is at the VBM ($\mu_e=0$). This $\epsilon(q_1/q_2)$ level [also referred to as the $(q_1/q_2)$ level], corresponding to a {\it defect level}, would be observed in, e.g., deep-level transient spectroscopy (DLTS) experiments where the defect in the final charge state $q_2$ fully relaxes to its equilibrium configuration after the transition. It is evident from Eq.~(\ref{eq;tl}) that $\epsilon(q_1/q_2)$ is independent of the atomic chemical potentials.

Note that we also carry out computational tests using the 2$\times$2$\times$2 Monkhorst-Pack \textbf{k}-point mesh~\cite{monkhorst-pack}, specifically for the Er$_{\rm Ga}$-2O$_{\rm As}$ defect complex. Compared to the results obtained in $\Gamma$-only calculations, the use of this \textbf{k}-point mesh has negligible effects on the defect's local structure and thermodynamic transition level. It changes the $\epsilon(+/0)$ level of Er$_{\rm Ga}$-2O$_{\rm As}$ by only 50 meV, which is within the error bar of our calculations (about 0.1 eV), and the O--O distance in the complex by 0.4\% or less. 

We employ the \textsc{nonrad} code \cite{Turiansky2021CPC} to calculate configuration coordinate diagrams for optical transitions and nonradiative carrier capture coefficients. The code implements the first-principles approach of Alkauskas et al.~\cite{Alkauskas2014PRB}. Effective electron and hole masses are calculated from \textsc{vasp} data by using the post-processing \textsc{vaspkit} package \cite{Wang2021CPC}; $m_{\rm e} = 0.089m_0$ and $m_{\rm h} = 0.345m_0$.

\section{Results and Discussion}\label{sec;results}

\begin{figure}
\vspace{0.2cm}
\includegraphics*[width=\linewidth]{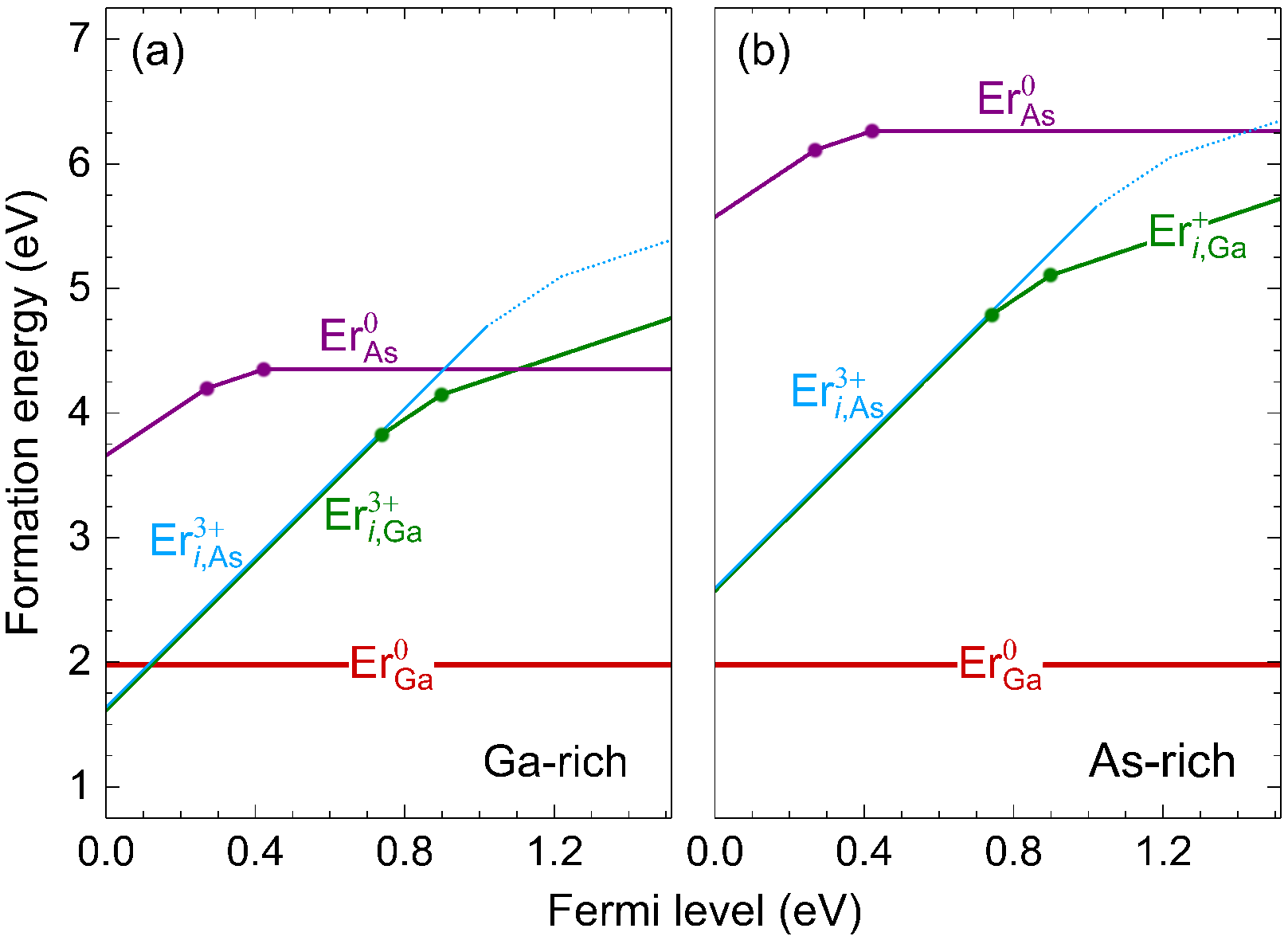}
\caption{Formation energies of isolated Er defects in GaAs, plotted as a function of the Fermi level from the VBM ($E_{\mathrm{v}}$, at 0 eV) to the CBM ($E_{\mathrm{c}}$, at 1.51 eV), under the extreme Ga-rich and As-rich conditions. For each defect, only segments corresponding to the lowest-energy charge states are shown. The slope indicates the charge state ($q$): positively (negatively) charged defects have positive (negative) slopes. Large solid dots connecting two segments mark the {\it defect levels}. The $2+$ and $+$ charge states of Er$_{i,{\rm As}}$ are unstable.}
\label{fig;isolated} 
\end{figure} 

\subsection{Isolated erbium defects}\label{sec;isolated}

Figure \ref{fig;isolated} shows the formation energy of isolated substitutional and interstitial Er defects in GaAs. The substitutional Er at the Ga lattice site (Er$_{\rm Ga}$) is stable only as electrically inactive Er$_{\rm Ga}^0$ (i.e., Er$^{3+}$ at the Ga site; spin $S = 3/2$). Note that in a defect notation, e.g., Er$_{\rm Ga}^0$, the superscript indicates the defect charge state, whereas in an ion notation, e.g., Er$^{3+}$, the superscript has the usual meaning (i.e., the oxidation state).  In this configuration, Er is tetrahedrally coordinated with As; the Er--As distance is 2.70--2.71 {\AA}, which is longer than the Ga--As bond (2.45 {\AA}) in the bulk, as Er$^{3+}$ has a larger ionic radius compared to Ga$^{3+}$~\cite{Shannon1976}; see Fig.~\ref{fig;struct}(a) in Appendix~\ref{sec;app}. In Er$_{i,{\rm Ga}}$, the interstitial Er atom is tetrahedrally coordinated with Ga atoms and octahedrally coordinated with As atoms (hereafter referred to as the interstitial site $T_{\rm Ga}$), see Figs.~\ref{fig;struct}(b) and ~\ref{fig;morestruct}(a), whereas in Er$_{i,{\rm As}}$ it is tetrahedrally coordinated with As and octahedrally coordinated with Ga (the $T_{\rm As}$ site), see Figs.~\ref{fig;struct}(c) and ~\ref{fig;morestruct}(b). In the substitutional Er at the As site (Er$_{\rm As}$), Er is significantly off-center [along the [11$\bar{1}$] direction by 1.74 {\AA} (in Er$_{\rm As}^0$) to 1.77{\AA} (Er$_{\rm As}^{2+}$)] and becomes coordinated with three Ga and three As atoms; see Figs.~\ref{fig;struct}(d) and ~\ref{fig;morestruct}(c) for the local structure of Er$_{\rm As}^{0}$. 

The structure of Er$_{\rm As}$ can be understood if we regard it not as a single defect but a defect complex. For example, Er$_{\rm As}^{0}$ ($S=3/2$) can be regarded as a complex of Er$_{i,{\rm As}}^{3+}$ ($S=3/2$) and the negatively charged As vacancy $V_{\rm As}^{3-}$ ($S=0$) with a binding energy of 3.48 eV with respect to its isolated constituents. In the $V_{\rm As}^{3-}$ constituent, one of the vacancy's adjacent Ga atoms is significantly displaced toward the vacancy (Such a displacement has also been observed in the isolated $V_{\rm As}^{3-}$ configuration~\cite{Hoang2026JPCM}). Table~\ref{tab;complex} presents stable charge and spin states of Er$_{\rm As}$ as complexes of their isolated constituents. Spins associated with each defect complex configuration and its constituents are determined based on an analysis of the calculated total and local magnetic moments of the defect complex as well as of its isolated constituents.    

In all these defect configurations, Er is stable only as the trivalent Er$^{3+}$. Note that the $2+$ and $+$ charge states of Er$_{i,{\rm As}}$ (the dotted lines in Fig.~\ref{fig;isolated}) are electronically unstable; i.e., the $(3+/2+)$ and $(2+/+)$ levels of Er$_{i,{\rm As}}$ are thus unstable. In this case, the addition of an electron to, e.g., the isolated Er$_{i,{\rm As}}^{3+}$ configuration does not result in Er$_{i,{\rm As}}^{2+}$ but Er$_{i,{\rm As}}^{3+}$ and a delocalized electron at the CBM in which the electron slightly perturbs the host conduction band. The neutral charge state of Er$_{i,{\rm As}}$ and Er$_{i,{\rm Ga}}$ is also unstable. The difference between Er$_{i,{\rm Ga}}$ and Er$_{i,{\rm As}}$ is rooted in their different lattice environments in which the local atomic (and electronic) structure of Er$_{i,{\rm As}}^{3+}$ (i.e., Er$^{3+}$ at the interstitial site $T_{\rm As}$) can be regarded as being equivalent to that of Er$_{\rm Ga}^0$ (i.e., Er$^{3+}$ at the Ga site).

Energetically, Er$_{\rm Ga}^0$ is most favorable among the isolated Er defects, except under the extreme Ga-rich condition where Er$_{i,{\rm Ga}}$ and Er$_{i,{\rm As}}$ have the lowest energy in a small range of Fermi-level values near the VBM, i.e., under p-type conditions; see Fig.~\ref{fig;isolated}(a). As one moves away from the extreme Ga-rich condition, the Er interstitials have a higher energy than Er$_{\rm Ga}$ even at the VBM. Er$_{i,{\rm Ga}}$ and Er$_{i,{\rm As}}$ would thus be less favorable than Er$_{\rm Ga}$ under realistic conditions. They are almost degenerate in energy in the $3+$ charge state (As shown in Sec.~\ref{sec;oxygen}, these two defects become more distinct when codoped with oxygen where Er$_{i,{\rm As}}$-O$_{\rm As}$ has a lower formation energy than Er$_{i,{\rm Ga}}$-O$_{\rm As}$. In other words, Er$_{i,{\rm As}}$ would be more favorable than Er$_{i,{\rm Ga}}$ when incorporated together with oxygen. As also seen later, the $2+$ and $+$ charge states of Er$_{i,{\rm As}}$ become stable in the presence of O$_{\rm As}$). Er$_{\rm As}$ is, however, much higher in energy than the other isolated Er defects and thus unlikely to form. Er$_{i,{\rm Ga}}$ and Er$_{\rm As}$ induce defect levels in the host band gap region, see also Table~\ref{tab;defectlevel}, and are thus electrically active. 

Our results are thus consistent with the metastability of Er interstitials observed in Er-implanted GaAs samples and the loss of the optical activity under host photoexcitation as Er$^{3+}$ ions move from the interstitial (i.e., Er$_{i,{\rm Ga}}$) to substitutional (Er$_{\rm Ga}$) sites after thermal annealing~\cite{Kozanecki1991SSC}. A detailed understanding of such a movement and of the effects of thermal annealing in general is, however, beyond the scope of the current work. 

For comparison, Taguchi and Ohno~\cite{Taguchi1997PRB} identified Er$_{\rm Ga}^0$ (i.e., Er$^{3+}$) as the most stable charge state in GaAs, except near the band edges. Their results showing the $(+/0)$ level near the VBM and the $(0/-)$ level near the CBM and the claim that ``Er$_{\rm Ga}$ acts as both a shallow donor and a shallow acceptor'' are not supported by our work. Svane et al.~\cite{Svane2006} identified Er$^{3+}$ as the ground-state configuration of the substitutional Er impurity in GaAs and found ``the divalent acceptor level'' $(0/-)$ above the CBM. We, on the other hand, find that the divalent Er$^{2+}$ is not stable anywhere, even above the CBM.

\begin{figure}
\vspace{0.2cm}
\includegraphics*[width=\linewidth]{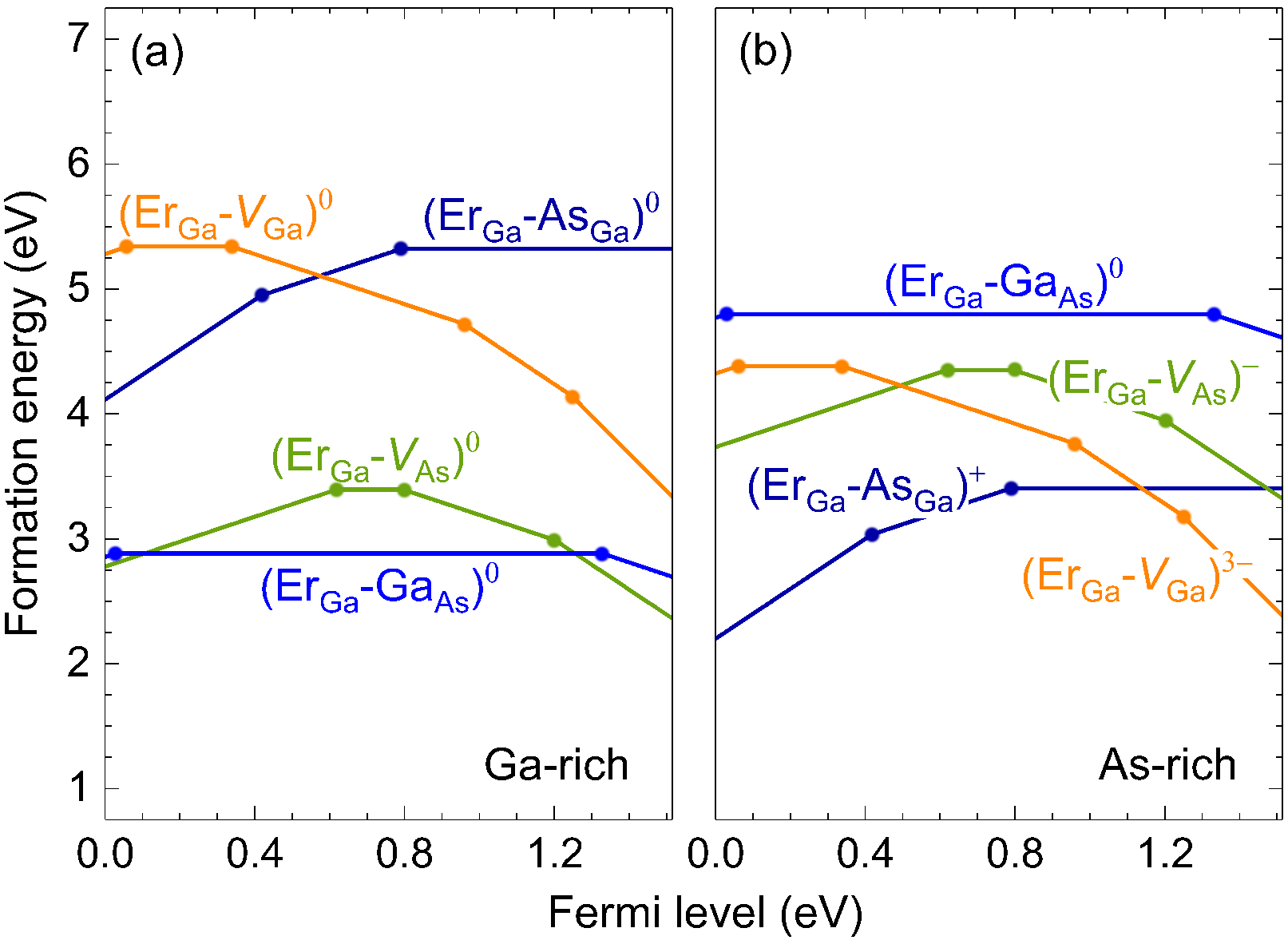}
\caption{Formation energies of complexes of Er and native defects under the extreme Ga-rich and As-rich conditions.}
\label{fig;natives} 
\end{figure}

\subsection{Erbium-native defect complexes}\label{sec;native}

Figure \ref{fig;natives} shows the formation energy of complexes consisting of Er$_{\rm Ga}$ and a native point defect (i.e., the antisite defect Ga$_{\rm As}$ or As$_{\rm Ga}$, or the vacancy $V_{\rm Ga}$ or $V_{\rm As}$). A detailed investigation of the native defect constituents has already been reported in Ref.~\citenum{Hoang2026JPCM}. Figures~\ref{fig;struct}(e)--\ref{fig;struct}(h) show the local structure of select defect complex configurations. In Er$_{\rm Ga}$-Ga$_{\rm As}$, both the constituents are significantly off-center where Er is three-fold coordinated with As and Ga is three-fold coordinated with Ga; Fig.~\ref{fig;struct}(e). This behavior is due to (i) the tendency of Ga$_{\rm As}$ to be off-center~\cite{Hoang2026JPCM} and (ii) the larger ionic radius of Er$^{3+}$ (compared to Ga$^{3+}$) which results in the longer Er--As bond (compared to Ga--As) and in Er being displaced away from the adjacent As atoms and toward the off-center Ga$_{\rm As}$. In Er$_{\rm Ga}$-As$_{\rm Ga}$, both the constituents are four-fold coordinated and on-center, Fig.~\ref{fig;struct}(f), consistent with the stable configuration of As$_{\rm Ga}$ being on-center~\cite{Hoang2026JPCM} (The metastable configuration of As$_{\rm Ga}$~\cite{Hoang2026JPCM}, often denoted as As$_{\rm Ga}^\ast$ in the literature, is thus not stable in the presence of Er$_{\rm Ga}$). In Er$_{\rm Ga}$-$V_{\rm Ga}$ and Er$_{\rm Ga}$-$V_{\rm As}$, Er is off-center and relaxes slightly toward the vacancy; Figs.~\ref{fig;struct}(g) and \ref{fig;struct}(h). We report in Table~\ref{tab;complex} the decomposition of the complex configurations into their isolated constituents.

All these Er-native defect complexes induce defect energy levels in the host band gap and are thus electrically active; see also Table~\ref{tab;defectlevel}. This property is largely inherited from the electrical activity of the native defect constituents~\cite{Hoang2026JPCM}. The binding energy of the complexes with respect to their isolated constituents is small, however, see Table~\ref{tab;complex}; and thus the complexes are unlikely to be stable and might occur only under thermodynamic non-equilibrium. As discussed in Ref.~\citenum{walle:3851}, in order for a defect complex to have a higher concentration than its isolated constituents, the binding energy needs to be greater than the larger of the formation energies of the isolated constituents. Although they may still form under non-equilibrium conditions where the constituents can be trapped together, we are not yet aware of any experimental report of any of these complexes. 

\subsection{Erbium-oxygen complexes}\label{sec;oxygen}

\begin{figure}
\vspace{0.2cm}
\includegraphics*[width=\linewidth]{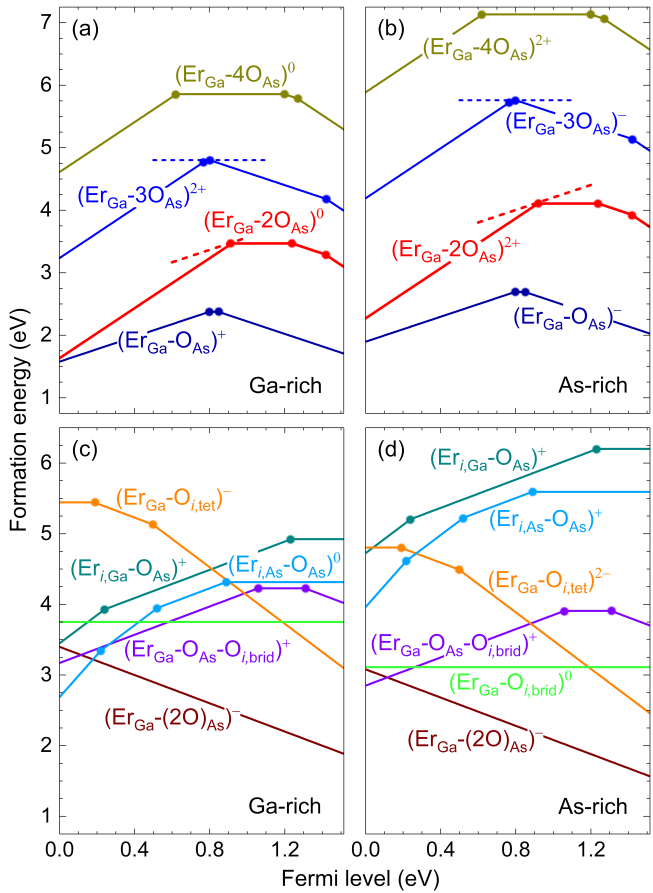}
\caption{Formation energies of various (Er,O)-related defect complexes in GaAs, under the extreme Ga-rich and As-rich conditions. The dashed segments represent energetically metastable (but electronically stable) charge states.}
\label{fig;oxygen} 
\end{figure}

Figure \ref{fig;oxygen} shows the formation energy of various defect complexes consisting of a Er-related defect (Er$_{\rm Ga}$, Er$_{i,{\rm Ga}}$, or Er$_{i,{\rm As}}$) and one or more O-related defects [specifically, O$_{\rm As}$, (2O)$_{\rm As}$, O$_{i, {\rm brid}}$, and/or O$_{i, {\rm tet}}$]. O$_{\rm As}$ is the substitutional O impurity at the As lattice site; (2O)$_{\rm As}$ is a configuration in which an As is substituted by two O atoms; O$_{i, {\rm brid}}$ is the interstitial O that bridges a Ga and an As and forms a puckered bond-center Ga--O--As structure; and O$_{i, {\rm tet}}$ is the interstitial O that is tetrahedrally coordinated with Ga. Details of the structural, electronic, and optical properties of the isolated O-related defects can be found in Ref.~\citenum{Hoang2026JPCM} in which the calculations were carried out using the same set of computational methods and procedures as in this work. In the (Er,O)-related complexes, the identities of the constituent defects are largely preserved but both Er and O are significantly off-center, see Figs.~\ref{fig;struct}(i)--\ref{fig;struct}(p) and \ref{fig;morestruct}(d), including those in Er$_{\rm Ga}$-4O$_{\rm As}$. The off-centering is mainly due to the shorter Er--O bond (and the longer Er--As bond) compared to the Ga--As bond. In (Er$_{\rm Ga}$-2O$_{\rm As}$)$^{2+}$, the most stable configuration of Er$_{\rm Ga}$-2O$_{\rm As}$ under p-type conditions and can be identified as the Er-2O center due to its $C_{2v}$ symmetry, the Er--O distance (and the Er--As distance) is 2.15 {\AA} (2.84 {\AA}), see Fig.~\ref{fig;struct}(j), which is comparable to that of 2.09--2.14 {\AA} (2.76--2.81 {\AA}) that Ofuchi et al.~\cite{Ofuchi2000ME} obtained in an extended x-ray absorption fine structure (EXAFS) analysis of (Er,O)-doped GaAs samples in which the majority of Er atoms reportedly formed the Er-2O center. In the charge states from $+$ to $2-$, Er$_{\rm Ga}$-2O$_{\rm As}$ deviates from the $C_{2v}$ symmetry as the two Er--O bonds are no longer identical; e.g., 2.14 {\AA} and 2.17 {\AA} in the $+$ state or 2.14 {\AA} and 2.24 {\AA} in the neutral state. The decomposition of the Er-related defect complex configurations into their isolated constituents is reported in Table~\ref{tab;complex}.

In Er$_{\rm Ga}$-$n$O$_{\rm As}$ ($n$ = 1--4), the defects all introduce energy levels in the host band gap, see Table~\ref{tab;defectlevel}. The defect levels induced by Er$_{\rm Ga}$-O$_{\rm As}$ largely resemble those of the isolated O$_{\rm As}$~\cite{Hoang2026JPCM}, whereas those by the other Er$_{\rm Ga}$-$n$O$_{\rm As}$ complexes in the series are significantly different due to the O$_{\rm As}$--O$_{\rm As}$ interaction. Among the other (Er,O)-related complexes, Er$_{\rm Ga}$-O$_{i,{\rm brid}}$ is stable only as (Er$_{\rm Ga}$-O$_{i,{\rm brid}}$)$^0$, which inherits the electrical inactivity of both the isolated Er$_{\rm Ga}$ and O$_{i,{\rm brid}}$~\cite{Hoang2026JPCM}. Er$_{\rm Ga}$-(2O)$_{\rm As}$ is stable only in the $-$ state, i.e., a shallow acceptor, determined solely by the isolated (2O)$_{\rm As}$~\cite{Hoang2026JPCM}. In Er$_{\rm Ga}$-O$_{i,{\rm tet}}$, Er$_{\rm Ga}$-O$_{\rm As}$-O$_{i,{\rm brid}}$, Er$_{i,{\rm Ga}}$-O$_{\rm As}$, and Er$_{i,{\rm As}}$-O$_{\rm As}$, the electrical activity of the complex also comes from that of the O constituent(s). Notably, the $(2+/+)$ and $(+/0)$ levels of Er$_{\rm Ga}$-2O$_{\rm As}$ are at 0.93 eV and 0.90 eV above the VBM (see Table~\ref{tab;defectlevel}), respectively, which are within the 0.82--1.22 region deduced by Hogg et al.~\cite{Hogg1996JAP}. The $(0/-)$ level at 1.24 eV (or $E_{\rm c}-0.27$ eV) appears to be closer to the estimated trap level at about 0.4 eV reported in the literature~\cite{Taguchi1996JAP,Takahei1996MRS,Hogg1997PRB}; however, the reliability of this estimated value depends on that of the model used in the fitting of experimental data. Besides, as discussed above, only the (Er$_{\rm Ga}$-2O$_{\rm As}$)$^{2+}$ configuration possesses the $C_{2v}$ symmetry. Note also that, as seen in Table~\ref{tab;defectlevel}, Er$_{\rm Ga}$-2O$_{\rm As}$ is not the only defect with energy levels in that range. Further computational characterization, as presented in Sec.~\ref{sec;nonrad}, is needed to have a more definite identification of the most efficient Er luminescence center.   

Figure~\ref{fig;compared} shows Er$_{\rm Ga}$-2O$_{\rm As}$ and the lowest-energy Er-related defects in GaAs, under either the extreme Ga-rich or As-rich condition, for easy comparison. We find that (Er,O)-related complexes can have an even lower formation energy than the isolated Er$_{\rm Ga}$. This indicates that codoping with oxygen facilitates the incorporation of Er into the GaAs host. Oxygen has also been known to stabilize Er in GaAs and eliminate the formation of ErAs precipitates~\cite{Maltez2004NIMPR}. The binding energy of the defect complexes with respect to their isolated constituents are large, which is consistent with the known strong affinity of rare earths to oxygen; see Table~\ref{tab;complex}. We also find that defect complexes such as Er$_{\rm Ga}$-2O$_{\rm As}$ form most easily under p-type (and Ga-rich) conditions; see Figs.~\ref{fig;oxygen} and \ref{fig;compared}. In (Er,O)-doped GaAs, Er$_{\rm Ga}$ might be incorporated first as Er$_{\rm Ga}$-O$_{\rm As}$, and the complex can then capture an O$_{\rm As}$ to form Er$_{\rm Ga}$-2O$_{\rm As}$. Under n-type conditions, Er$_{\rm Ga}$-O$_{\rm As}$ and Er$_{\rm Ga}$-(2O)$_{\rm As}$ are energetically most favorable under the Ga- and As-rich conditions, respectively, whereas the formation energy of Er$_{\rm Ga}$-2O$_{\rm As}$ is much higher. Our results thus clearly explain why the formation of the Er-2O center is greatly suppressed in n-type doping~\cite{Fujiwara2005MT}.

The formation energy of Er$_{i,{\rm As}}$-O$_{\rm As}$ is lower than that of Er$_{i,{\rm Ga}}$-O$_{\rm As}$, indicating that it is energetically more favorable to incorporate Er into GaAs at the interstitial site $T_{\rm As}$ than at the $T_{\rm Ga}$ site when codoped with oxygen. The $2+$ and $+$ charge states of Er$_{i,{\rm As}}$, unstable when Er$_{i,{\rm As}}$ is in isolation, become stable in the Er$_{i,{\rm As}}$-O$_{\rm As}$ complex due to the distorted lattice environment created by the presence of O$_{\rm As}$. Still, Er$_{i,{\rm Ga}}$-O$_{\rm As}$ and Er$_{i,{\rm As}}$-O$_{\rm As}$ are much higher in energy than, e.g., Er$_{\rm Ga}$, Er$_{\rm Ga}$-O$_{\rm As}$, Er$_{\rm Ga}$-2O$_{\rm As}$, and Er$_{\rm Ga}$-(2O)$_{\rm As}$, which can explain why Er is mostly at the near-substitutional sites in (Er,O)-doped GaAs samples~\cite{Takahei1994JAP,Kaczanowski1996NIM,Ofuchi2000ME,Takahei1997JAP}. Note that Er$_{\rm Ga}$-4O$_{\rm As}$ is also very high in energy and unlikely to occur with a sizable concentration under thermodynamic equilibrium; see Fig.~\ref{fig;oxygen}. 

For comparison, Coutinho et al.~\cite{Coutinho2004APL} reported results for Er$_{\rm Ga}$O$_{\rm As}$ [which is Er$_{\rm Ga}$-O$_{\rm As}$ in our notation], Er$_{\rm Ga}$O$_{i}$ [Er$_{\rm Ga}$-O$_{i,{\rm brid}}$], Er$_{\rm Ga}$(O$_{\rm As}$)$_2$ [Er$_{\rm Ga}$-2O$_{\rm As}$], Er$_{\rm Ga}$O$_{\rm As}$O$_i$(A) [Er$_{\rm Ga}$-(2O)$_{\rm As}$], and Er$_{\rm Ga}$O$_{\rm As}$O$_i$(B) [Er$_{\rm Ga}$-O$_{\rm As}$-O$_{i,{\rm brid}}$]. Among these (Er,O)-related defect complexes, the results for Er$_{\rm Ga}$O$_{\rm As}$, Er$_{\rm Ga}$(O$_{\rm As}$)$_2$, and Er$_{\rm Ga}$O$_{\rm As}$O$_i$(B) are in qualitative agreement with our results for Er$_{\rm Ga}$-O$_{\rm As}$, Er$_{\rm Ga}$-2O$_{\rm As}$, and Er$_{\rm Ga}$-O$_{\rm As}$-O$_{i,{\rm brid}}$, respectively. Their calculated $(0/-)$ level of Er$_{\rm Ga}$(O$_{\rm As}$)$_2$ at $E_{\rm c}-0.21$ eV has been considered as being in fair agreement with the trap level at about 0.4 eV reported in the literature~\cite{Taguchi1996JAP,Takahei1996MRS,Hogg1997PRB}. The authors, however, did not report on the $(2+/+)$ level of Er$_{\rm Ga}$(O$_{\rm As}$)$_2$. As discussed below, we find that the most efficient trap level is not the $(0/-)$ level but the $(2+/+)$ level, and the $(0/-)$ trap level associated with Er$_{\rm Ga}$-2O$_{\rm As}$ is even less efficient than that associated with Er$_{\rm Ga}$-O$_{\rm As}$. Finally, they found that Er$_{\rm Ga}$O$_{i}$ induces the $(+/0)$ level at $E_{\rm v}+0.06$ eV and Er$_{\rm Ga}$O$_{\rm As}$O$_i$(A) induces the $(0/-)$ level at $E_{\rm c}-1.08$ eV, which is in contrast to our results which show that Er$_{\rm Ga}$-O$_{i,{\rm brid}}$ and Er$_{\rm Ga}$-(2O)$_{\rm As}$ do not induce any defect levels in the host band gap. 

\begin{figure}
\vspace{0.2cm}
\includegraphics*[width=\linewidth]{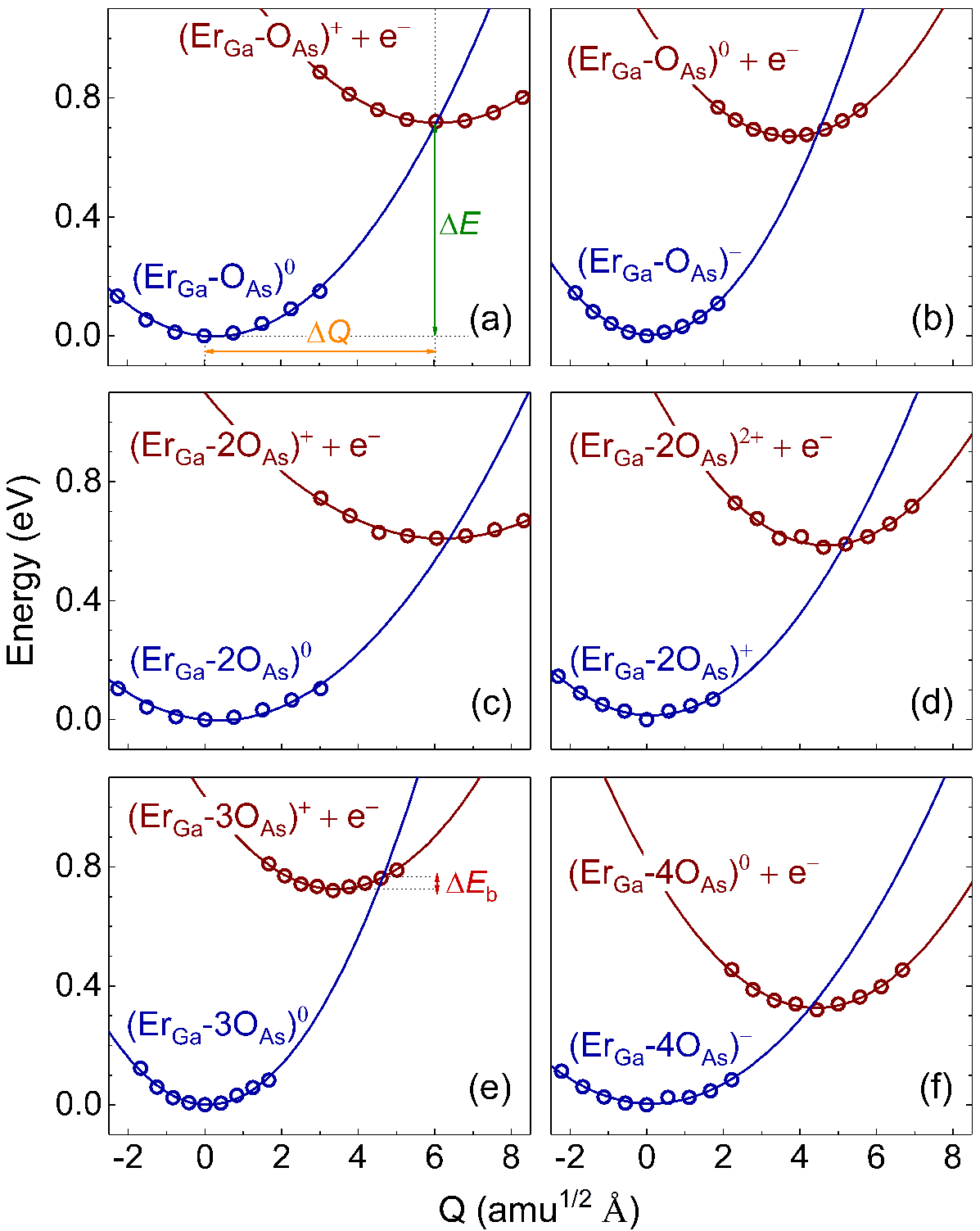}
\caption{Configuration coordinate diagrams for transitions involving select Er-related defect complexes in GaAs. $\Delta E$ is the ionization energy (trap depth); $\Delta Q$ is the mass-weighted difference between the geometries of the excited and ground states; and $\Delta E_b$ is the (semiclassical) electron capture barrier. See also Fig.~\ref{fig;morecc} for other transitions considered in this work.}
\label{fig;cc} 
\end{figure}

\subsection{Trap-assisted recombination centers}\label{sec;nonrad}

Our results thus far show that there are many possible Er-related defects in GaAs. The electronic structure and energetics of the defects are generally asymmetric with respect to n- and p-type doping; many have a lower formation energy under p-type conditions and defect levels that tend to favor electron trapping (i.e., with carrier-capturing configurations that are non-repulsive to electrons). Exceptions are Er$_{\rm Ga}$-$V_{\rm Ga}$ and Er$_{\rm Ga}$-$V_{\rm As}$, but these complexes have a very low binding energy and are unlikely to be stable; Er$_{\rm Ga}$-O$_{i,{\rm tet}}$, but it is much higher in energy than the lowest-energy (Er,O)-related defects; and Er$_{\rm Ga}$-(2O)$_{\rm As}$, which does not have any defect level in the host band gap and is thus optically inactive under host photoexcitation. We will thus focus on defects and defect levels that are suitable for electron trapping. As the recombination rate is limited by the rate of the capture of minority carriers, such levels would be efficient as trap-assisted recombination centers in p-type GaAs.

Among the energy levels induced by the isolated Er-related defects, only the $(2+/+)$ level of Er$_{i,{\rm Ga}}$, at 0.90 eV above the VBM, is suitable as a recombination center for Er$^{3+}$ excitation via an electron trapping mechanism. Among the Er-native defect complexes, the $(0/-)$ level of Er$_{\rm Ga}$-Ga$_{\rm As}$, at 1.33 eV, would also be suitable; however, the complex is likely unstable and the mismatch between the recombination energy (1.33 eV) and the intra-$4f$-excitation energy ($\sim$0.81 eV) is large. Most notably, we find that the (Er,O)-related defects offer many suitable defect levels (Table~\ref{tab;defectlevel}) in which the $(2+/+)$ level of Er$_{\rm Ga}$-2O$_{\rm As}$, at 0.93 eV, emerges as an outstanding candidate for Er$^{3+}$ excitation in p-type GaAs, based on the criteria discussed in Sec.~\ref{sec;intro}. The carrier-capturing configuration of this center, i.e., (Er$_{\rm Ga}$-2O$_{\rm As}$)$^{2+}$, has a low formation energy under p-type conditions, is highly attractive to the minority carriers (i.e., electrons from the conduction band), and has a small energy mismatch of $\sim$0.1 eV or even smaller (if one accounts for the shallow acceptor binding energy, about 30--40 meV for Be, Mg, or Zn at the Ga site or C, Si, or Ge at the As site~\cite{Fiorentini1995PRB}).

To quantify the ability of defect centers to capture charge carriers from the band states, we calculate their nonradiative carrier capture coefficients~\cite{Alkauskas2014PRB,Turiansky2021CPC}. Specific defects include Er$_{i,\rm Ga}$, Er$_{\rm Ga}$-Ga$_{\rm As}$, Er$_{\rm Ga}$-As$_{\rm Ga}$, Er$_{\rm Ga}$-$n$O$_{\rm As}$ ($n$ = 1--4), and Er$_{\rm Ga}$-O$_{\rm As}$-O$_{i,{\rm brid}}$. Some of the transition levels associated with these defects do not satisfy the condition $\epsilon(q_1/q_2) > 0.81$ eV and some defect complexes have low binding energies but are included for comparison and to see the effect of defect association. Note that although the calculation of nonradiative carrier capture coefficients (and cross sections) following the first-principles approach of Alkauskas et al.~\cite{Alkauskas2014PRB,Turiansky2021CPC} has been shown to be reliable and generally in good agreement with experiments (at least within an order of magnitude)~\cite{Alkauskas2014PRB,Hoang2026JPCM}, in the following we focus mainly on the trend among the RE-related defects and their trap levels.

Figures~\ref{fig;cc} and \ref{fig;morecc} show the configuration coordinate diagram for select transitions, obtained using the \textsc{nonrad} code~\cite{Turiansky2021CPC}, which expresses the total energy as a function of the generalized coordinate $Q$ which captures collective atomic displacements in an one-dimensional approximation by linear interpolation between geometries of excited and ground states~\cite{Alkauskas2014PRB}. Table~\ref{tab;opt} lists the parameters that appear in the nonradiative capture coefficient~\cite{Alkauskas2014PRB,Turiansky2021CPC}. These include the ionization energy or the trap depth ($\Delta E$) which, for electron capture, is the thermodynamic transition level $\epsilon(q_1/q_2)$ referenced to the CBM, and key parameters extracted from \textsc{nonrad} calculations, including the mass-weighted difference ($\Delta Q$) between the geometries of the excited (e) and ground (g) states, configurational degeneracy at the ground state ($g_g$) which is the number configurations that the carrier can be captured into, energies of the effective vibrations ($\hbar \Omega_{\{\rm e,g\}}$), and electron-phonon coupling matrix elements ($\tilde{W}_{\rm eg}$).

\begin{figure}
\vspace{0.2cm}
\includegraphics*[width=\linewidth]{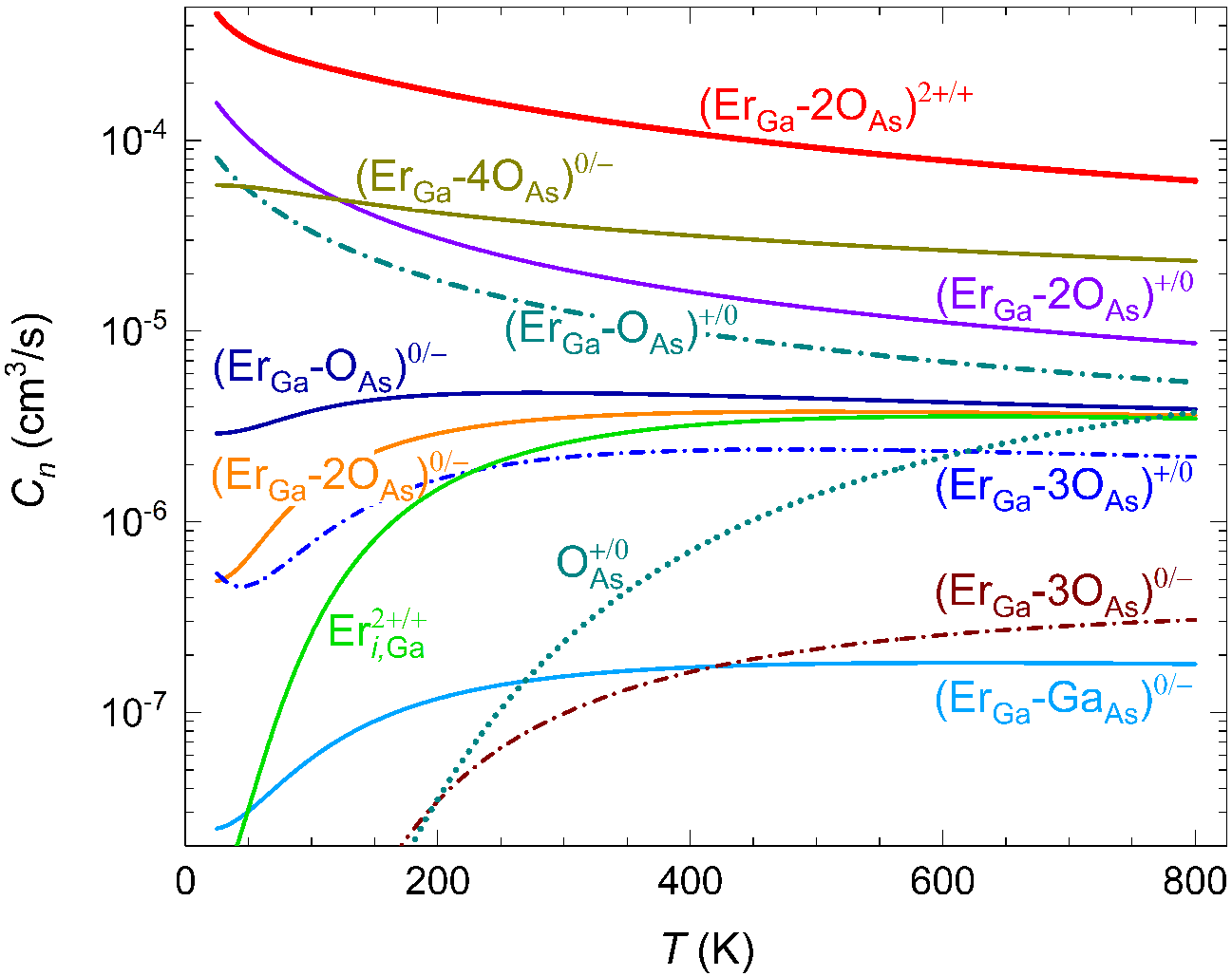} 
\caption{Nonradiative electron capture coefficient ($C_n$) at different Er-related defect centers in GaAs. Solid (dash-dotted) curves correspond to transitions with nonradiative recombination energies greater (smaller) than $0.81$ eV. The result for O$_{\rm As}$ (dotted curve) is also included for comparison.}
\label{fig;Cn} 
\end{figure} 

Figure~\ref{fig;Cn} shows the nonradiative electron capture coefficients ($C_n$). At the X$^{q_1/q_2}$ defect center associated with each $C_n$ curve, X$^{q_1}$ is the configuration into which an electron is captured. Solid $C_n$ curves in the figure are those with $\epsilon(q_1/q_2)>0.81$ eV. Among the defects under consideration, we find that (Er$_{\rm Ga}$-2O$_{\rm As}$)$^{2+/+}$ is the most efficient center for electron capture. The highest $C_n$ value associated with this center is resulted from having an almost zero semiclassical electron capture barrier [i.e., $\Delta E_b \sim 0$ eV, see Fig.~\ref{fig;cc}(d)], a large Sommerfeld parameter (due to the long-range Coulombic attraction between the electron and the doubly positively charged carrier-capturing configuration), and a relatively large electron-phonon coupling matrix element. (Er$_{\rm Ga}$-4O$_{\rm As}$)$^{0/-}$ also has a high capture coefficient, but it has a larger energy mismatch ($\sim$0.3 eV) than (Er$_{\rm Ga}$-2O$_{\rm As}$)$^{2+/+}$. Notably, we find that Er$_{i,\rm Ga}^{2+/+}$ is as efficient as some of the (Er,O)-related centers, which can explain the Er luminescence observed in pre-annealed, Er-implanted GaAs samples~\cite{Kozanecki1991SSC}. (Er$_{\rm Ga}$-Ga$_{\rm As}$)$^{0/-}$ also has a sizable electron capture coefficient value but is less efficient than most of the (Er,O)-related centers. Other centers such as (Er$_{\rm Ga}$-As$_{\rm Ga}$)$^{+/0}$ and (Er$_{\rm Ga}$-O$_{\rm As}$-O$_{i,{\rm brid}}$)$^{+/0}$ has negligible capture coefficients (and not included in Fig.~\ref{fig;Cn}) mainly due to the high electron capture barrier; $\Delta E_b \sim 1.5$ eV [in the former, Fig.~\ref{fig;morecc}(c)] or $\infty$ [in the latter, Fig.~\ref{fig;morecc}(d)]. The trend in going from Er$_{\rm Ga}$-As$_{\rm Ga}$ to Er$_{\rm Ga}$-Ga$_{\rm As}$ to Er$_{\rm Ga}$-O$_{\rm As}$ follows that of the isolated As$_{\rm Ga}$ to Ga$_{\rm As}$ to O$_{\rm As}$~\cite{Hoang2026JPCM}. But the electron capture coefficient at (Er$_{\rm Ga}$-O$_{\rm As}$)$^{+}$ is about 54 times higher than that at the isolated O$_{\rm As}^+$ at room temperature, which can be ascribed in a large part to the higher semiclassical capture barrier in the latter case [$\Delta E_b \sim 0.3$ eV, Fig.~\ref{fig;morecc}(f), compared to $\sim$ 0 eV in the former, Fig.~\ref{fig;cc}(a)]. As also seen in Fig.~\ref{fig;Cn}, the interaction between O-related constituents in the (Er,O)-related complexes has a significant impact on nonradiative carrier capture. Note that dash-dotted $C_n$ curves in Fig.~\ref{fig;Cn} are those with $\epsilon(q_1/q_2)<0.81$ eV, i.e., centers with low recombination energies and thus optically inactive under host photoexcitation, but they can be excited under direct intra-$4f$-shell photoexcitation. 

Although the capture of majority carriers is not a concern due to the high carrier concentration, we nevertheless calculate the nonradiative hole capture coefficient ($C_p$) at relevant Er$_{\rm Ga}$-2O$_{\rm As}$ and Er$_{\rm Ga}$-4O$_{\rm As}$ configurations for comparison. Figure~\ref{fig;Cp} shows that (Er$_{\rm Ga}$-2O$_{\rm As}$)$^{+}$ is more efficient than (Er$_{\rm Ga}$-4O$_{\rm As}$)$^{-}$ at capturing a hole, mainly due to the higher electron-phonon coupling matrix element in the former (more than three times higher than in the latter). Combined with the above results, this again indicates that, among the Er-related defects, (Er$_{\rm Ga}$-2O$_{\rm As}$)$^{2+/+}$ is overall the most efficient trap-assisted recombination center for Er$^{3+}$ excitation, and can be identified with type-I defects discussed in Sec.~\ref{sec;intro}. Type-II defects are those without a defect level in the host band gap [e.g., Er$_{\rm Ga}$ or Er$_{\rm Ga}$-(2O)$_{\rm As}$] or those centers with an insufficient recombination energy for Er$^{3+}$ excitation. Note that here we do not yet consider defect complexes with direct Er--Er interaction or ``aggregates of Er complexes'' that Takahei and Taguchi~\cite{Takahei1995JAP} appeared to allude to in their discussion of type-III defects.    

Finally, regarding the effects of the Er/O ratio on the Er PL, Maltez et al.~\cite{Maltez2004NIMPR} found that the Er:2O ratio is optimal, whereas excessive oxygen leads to performance degradation. This appears to be consistent with our results showing that in going from Er$_{\rm Ga}$-2O$_{\rm As}$ to Er$_{\rm Ga}$-3O$_{\rm As}$ the defect loses its optical activity under host photoexcitation. The reason is not just because the nonradiative capture coefficient at Er$_{\rm Ga}$-3O$_{\rm As}$ is lower (as seen in Fig.~\ref{fig;Cn}) but, more importantly, the recombination energy at this defect becomes insufficient for Er$^{3+}$ excitation (Fig.~\ref{fig;oxygen}). EXAFS analysis~\cite{Ofuchi2000ME} appears to suggest the presence of the optically inefficient Er$_{\rm Ga}$-3O$_{\rm As}$ defect in (Er,O)-doped GaAs samples with a high Er concentration and a high O (low As) coordination number.  

\section{Conclusions} 

We have carried out a systematic investigation of Er-related defects in GaAs using hybrid functional calculations. The defect electronic structure and energetics are found to be generally asymmetric with respect to n- and p-type doping, with many of the defects have a lower formation energy under p-type conditions and defect levels that tend to favor electron trapping. On the basis of the calculated defect levels, formation energies, and nonradiative carrier capture coefficients, we identify Er-related defect centers that are optically active under host photoexcitation or via minority carrier injection. The (2+/+) transition level of the defect complex Er$_{\rm Ga}$-2O$_{\rm As}$, in which the electron-capturing configuration (Er$_{\rm Ga}$-2O$_{\rm As}$)$^{2+}$ possesses the $C_{2v}$ symmetry and can be incorporated into GaAs more easily under p-type conditions, is found to be the most efficient trap-assisted recombination center for Er$^{3+}$ excitation. The Er-2O center observed in experiments can now be identified explicitly with the (Er$_{\rm Ga}$-2O$_{\rm As}$)$^{2+}$ configuration. Our results thus provide an understanding for why the Er-2O center is most efficient and for the effects of n- and p-type doping and of the Er/O ratio on the formation of optically active Er centers and on the Er luminescence observed in experiments. They offer guidelines for defect-controlled synthesis and further defect characterization. More importantly, the computational approach demonstrated here, which involves a combination of computational structural, electronic, and optical characterizations, can be applied to other RE-doped semiconductors and insulators and is efficient in screening RE defects for optoelectronics and quantum information applications.

\begin{acknowledgments}

The author is grateful to Mark Turiansky for helpful discussion. This work used resources of the Center for Computationally Assisted Science and Technology (CCAST) at North Dakota State University, which were made possible in part by National Science Foundation MRI Award No.~2019077.

\end{acknowledgments}

\section*{Data Availability}

The data that support the findings of this study are included in this article and its appendix. Additional data are available from the author upon reasonable request.

\appendix

\section{Supporting Material}\label{sec;app}

Additional information on defect levels and local structures; formation energies, binding energies, and other details of the defect complexes; configuration coordinate diagrams; and nonradiative hole capture coefficients. 

\renewcommand{\thefigure}{S1}
\begin{figure*}
\centering
\includegraphics[width=0.90\linewidth]{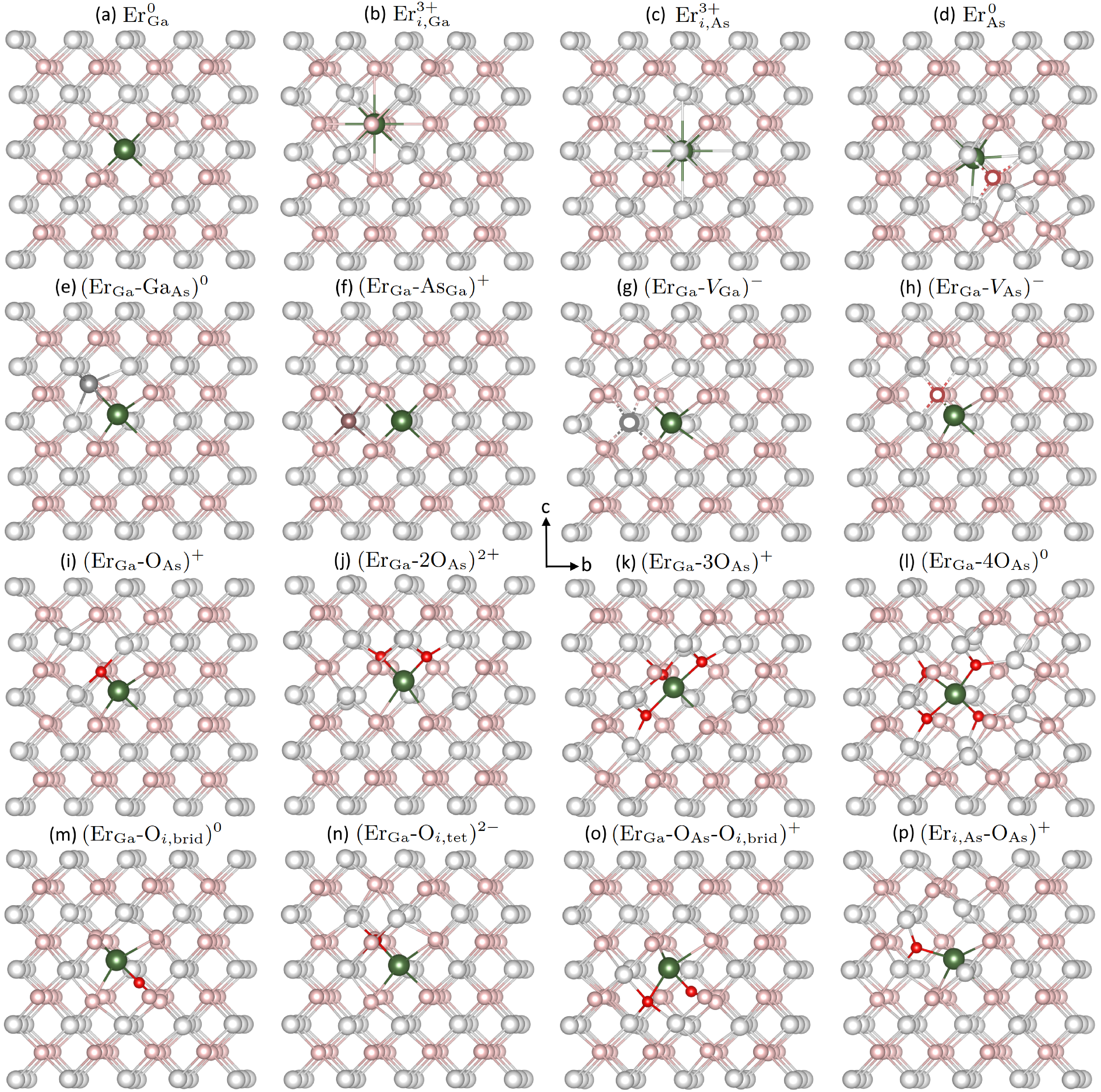}
\caption{Local structures of select Er-related defect configurations in GaAs. Large (gray) spheres are Ga and small (red) spheres are As where antisite Ga and As atoms are marked with slightly darker colors for easy recognition; oxygen is represented by a smaller (red) sphere, and the Ga (As) vacancy by a gray (red) empty circle. See also Fig.~\ref{fig;morestruct} for the local structure of Er$_{i, {\rm Ga}}^{3+}$, Er$_{i, {\rm As}}^{3+}$, Er$_{\rm As}^{0}$, and (Er$_{i, {\rm As}}$-O$_{\rm As}$)$^+$ viewed from a different angle, and Table~\ref{tab;complex} for details on all the defect complexes.}
\label{fig;struct}
\end{figure*}

\renewcommand{\thefigure}{S2}
\begin{figure*}
\centering
\includegraphics[width=\linewidth]{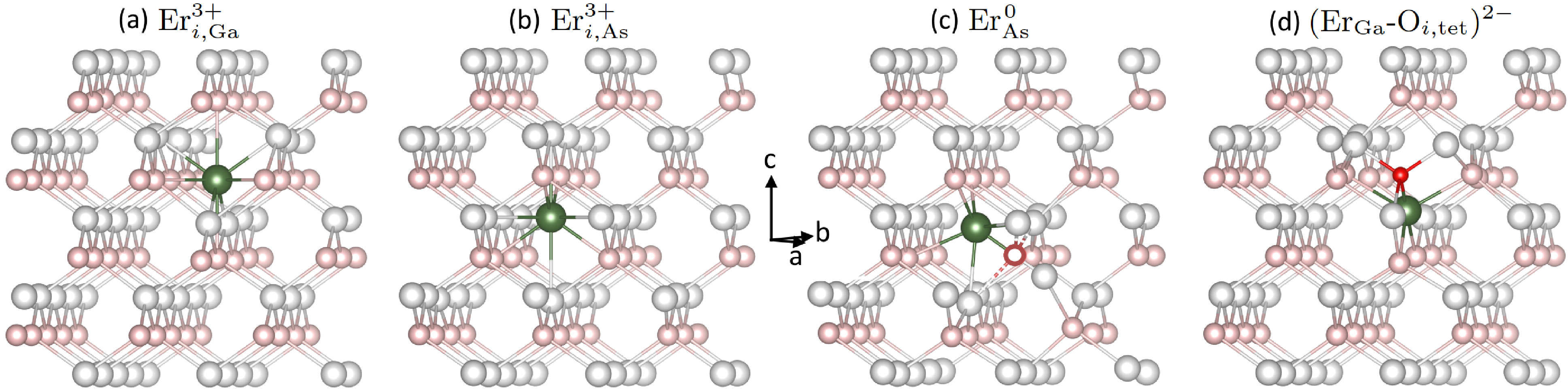}
\caption{Local structures of select Er-related defect configurations in GaAs viewed from a different angle. Er$_{\rm As}^{0}$ [in (c)] can be regarded as a complex of Er$_{i,{\rm As}}^{3+}$ [in (b)] and $V_{\rm As}^{3-}$; the As vacancy is represented by a red empty circle.}
\label{fig;morestruct}
\end{figure*}

\renewcommand\thetable{S1}
\begin{table*}
\caption{Stable charge and spin states of defect complexes, their constituent defects, and binding energies ($E_{\rm b}$).}\label{tab;complex}
\begin{ruledtabular}
\begin{tabular}{lclc}
Complex & Spin ($S$) & Constituents (Spin)$^a$ & $E_{\rm b}$ (eV)\\
\colrule
Er$_{\rm As}^{2+}$ &3/2& Er$_{i,{\rm As}}^{3+}$ ($S=3/2$) + $V_{\rm As}^-$ ($S=0$) & 1.60 \\
Er$_{\rm As}^{+}$ &2& Er$_{i,{\rm As}}^{3+}$ ($S=3/2$) + $V_{\rm As}^{2-}$ ($S=1/2$) & 2.94 \\
Er$_{\rm As}^{0}$ &3/2 & Er$_{i,{\rm As}}^{3+}$ ($S=3/2$) + $V_{\rm As}^{3-}$ ($S=0$) & 3.48 \\
(Er$_{\rm Ga}$-As$_{\rm Ga}$)$^{2+}$&$3/2$&Er$_{\rm Ga}^{0}$ ($S=3/2$) + As$_{\rm Ga}^{2+}$ ($S=0$)&0.04\\      
(Er$_{\rm Ga}$-As$_{\rm Ga}$)$^+$&2&Er$_{\rm Ga}^{0}$ ($S=3/2$) + As$_{\rm Ga}^{+}$ ($S=1/2$)&0.06\\
(Er$_{\rm Ga}$-As$_{\rm Ga}$)$^0$&$3/2$&Er$_{\rm Ga}^{0}$ ($S=3/2$) + As$_{\rm Ga}^{0}$ ($S=0$)&0.07\\
(Er$_{\rm Ga}$-Ga$_{\rm As}$)$^+$&2&Er$_{\rm Ga}^{0}$ ($S=3/2$) + Ga$_{\rm As}^{+}$ ($S=1/2$)&0.63\\
(Er$_{\rm Ga}$-Ga$_{\rm As}$)$^0$&$3/2$&Er$_{\rm Ga}^{0}$ ($S=3/2$) + Ga$_{\rm As}^{0}$ ($S=0$)&0.91\\
(Er$_{\rm Ga}$-Ga$_{\rm As}$)$^-$&2&Er$_{\rm Ga}^{0}$ ($S=3/2$) + Ga$_{\rm As}^{-}$ ($S=1/2$)&0.15\\
(Er$_{\rm Ga}$-$V_{\rm As}$)$^+$&$3/2$&Er$_{\rm Ga}^{0}$ ($S=3/2$) + $V_{\rm As}^{+}$ ($S=0$)&1.09\\
(Er$_{\rm Ga}$-$V_{\rm As}$)$^0$&2&Er$_{\rm Ga}^{0}$ ($S=3/2$) + $V_{\rm As}^{0}$ ($S=1/2$)&1.52\\
(Er$_{\rm Ga}$-$V_{\rm As}$)$^-$&$3/2$&Er$_{\rm Ga}^{0}$ ($S=3/2$) + $V_{\rm As}^{-}$ ($S=0$)&1.41\\
(Er$_{\rm Ga}$-$V_{\rm As}$)$^{2-}$&2&Er$_{\rm Ga}^{0}$ ($S=3/2$) + $V_{\rm As}^{2-}$ ($S=1/2$)&1.98\\
(Er$_{\rm Ga}$-$V_{\rm Ga}$)$^+$&$5/2$&Er$_{\rm Ga}^{0}$ ($S=3/2$) + $V_{\rm Ga}^{+}$ ($S=1$)&0.80\\
(Er$_{\rm Ga}$-$V_{\rm Ga}$)$^0$&2&Er$_{\rm Ga}^{0}$ ($S=3/2$) + $V_{\rm Ga}^{0}$ ($S=1/2$)&0.72\\
(Er$_{\rm Ga}$-$V_{\rm Ga}$)$^-$&$3/2$&Er$_{\rm Ga}^{0}$ ($S=3/2$) + $V_{\rm Ga}^{-}$ ($S=0$)&0.80\\
(Er$_{\rm Ga}$-$V_{\rm Ga}$)$^{2-}$&2&Er$_{\rm Ga}^{0}$ ($S=3/2$) + $V_{\rm Ga}^{2-}$ ($S=1/2$)&0.47\\
(Er$_{\rm Ga}$-$V_{\rm Ga}$)$^{3-}$&$3/2$&Er$_{\rm Ga}^{0}$ ($S=3/2$) + $V_{\rm Ga}^{3-}$ ($S=0$)& 0.11\\
(Er$_{\rm Ga}$-O$_{i,{\rm brid}}$)$^0$&$3/2$&Er$_{\rm Ga}^{0}$ ($S=3/2$) + O$_{i,{\rm brid}}^{0}$ ($S=0$)&0.78\\
(Er$_{\rm Ga}$-O$_{i,{\rm tet}}$)$^+$&2&Er$_{\rm Ga}^{0}$ ($S=3/2$) + O$_{i,{\rm tet}}^{+}$ ($S=1/2$)&1.26\\
(Er$_{\rm Ga}$-O$_{i,{\rm tet}}$)$^0$&$3/2$&Er$_{\rm Ga}^{0}$ ($S=3/2$) + O$_{i,{\rm tet}}^{0}$ ($S=0$)&1.28\\
(Er$_{\rm Ga}$-O$_{i,{\rm tet}}$)$^-$&2&Er$_{\rm Ga}^{0}$ ($S=3/2$) + O$_{i,{\rm tet}}^{-}$ ($S=1/2$)&1.27\\
(Er$_{\rm Ga}$-O$_{i,{\rm tet}}$)$^{2-}$&$3/2$&Er$_{\rm Ga}^{0}$ ($S=3/2$) + O$_{i,{\rm tet}}^{2-}$ ($S=0$)&1.07\\
(Er$_{\rm Ga}$-(2O)$_{\rm As}$)$^-$& $3/2$&Er$_{\rm Ga}^{0}$ ($S=3/2$) + (2O)$_{\rm As}^{-} ($S=0$)$&2.80\\
(Er$_{\rm Ga}$-O$_{\rm As}$-O$_{i,{\rm brid}}$)$^+$&$3/2$&Er$_{\rm Ga}^{0}$ ($S=3/2$) + O$_{\rm As}^{+}$ ($S=0$) + O$_{i,{\rm brid}}^0$ ($S=0$)&2.74\\
(Er$_{\rm Ga}$-O$_{\rm As}$-O$_{i,{\rm brid}}$)$^0$&2&Er$_{\rm Ga}^{0}$ ($S=3/2$) + O$_{\rm As}^{0}$ ($S=1/2$) + O$_{i,{\rm brid}}^0$ ($S=0$)&2.67\\
(Er$_{\rm Ga}$-O$_{\rm As}$-O$_{i,{\rm brid}}$)$^-$&$3/2$&Er$_{\rm Ga}^{0}$ ($S=3/2$) + O$_{\rm As}^{-}$ ($S=0$) + O$_{i,{\rm brid}}^0$ ($S=0$)&2.35\\
(Er$_{\rm Ga}$-O$_{\rm As}$)$^+$&$3/2$&Er$_{\rm Ga}^{0}$ ($S=3/2$) + O$_{\rm As}^{+}$ ($S=0$)&1.78\\
(Er$_{\rm Ga}$-O$_{\rm As}$)$^0$&$2$&Er$_{\rm Ga}^{0}$ ($S=3/2$) + O$_{\rm As}^{0}$ ($S=1/2$)&1.97\\
(Er$_{\rm Ga}$-O$_{\rm As}$)$^-$&$3/2$&Er$_{\rm Ga}^{0}$ ($S=3/2$) + O$_{\rm As}^{-}$ ($S=0$)&2.11\\
(Er$_{\rm Ga}$-2O$_{\rm As}$)$^{2+}$&$3/2$&Er$_{\rm Ga}^{0}$ ($S=3/2$) + O$_{\rm As}^{+}$ ($S=0$) + O$_{\rm As}^{+}$ ($S=0$)&3.10\\
(Er$_{\rm Ga}$-2O$_{\rm As}$)$^+$&2&Er$_{\rm Ga}^{0}$ ($S=3/2$) + O$_{\rm As}^{+}$ ($S=0$) + O$_{\rm As}^{0}$ ($S=1/2$)&3.15\\
(Er$_{\rm Ga}$-2O$_{\rm As}$)$^0$&$3/2$&Er$_{\rm Ga}^{0}$ ($S=3/2$) + O$_{\rm As}^{+}$ ($S=0$) + O$_{\rm As}^{-}$ ($S=0$)&3.24\\
(Er$_{\rm Ga}$-2O$_{\rm As}$)$^-$&2&Er$_{\rm Ga}^{0}$ ($S=3/2$) + O$_{\rm As}^{0}$ ($S=1/2$) + O$_{\rm As}^{-}$ ($S=0$)&2.99\\
(Er$_{\rm Ga}$-2O$_{\rm As}$)$^{2-}$&$3/2$&Er$_{\rm Ga}^{0}$ ($S=3/2$) + O$_{\rm As}^{-}$ ($S=0$) + O$_{\rm As}^{-}$ ($S=0$)&2.56\\
(Er$_{\rm Ga}$-3O$_{\rm As}$)$^{2+}$&2&Er$_{\rm Ga}^{0}$ ($S=3/2$) + 2O$_{\rm As}^{+}$ ($S=0$) + O$_{\rm As}^{0}$ ($S=1/2$)&3.86\\
(Er$_{\rm Ga}$-3O$_{\rm As}$)$^{+}$&$3/2$&Er$_{\rm Ga}^{0}$ ($S=3/2$) + 2O$_{\rm As}^{+}$ ($S=0$) + O$_{\rm As}^{-}$ ($S=0$)&4.08\\
(Er$_{\rm Ga}$-3O$_{\rm As}$)$^{0}$&2&Er$_{\rm Ga}^{0}$ ($S=3/2$) + O$_{\rm As}^{+}$ ($S=0$) + O$_{\rm As}^{-}$ ($S=0$) + O$_{\rm As}^{0}$ ($S=1/2$)&4.27\\
(Er$_{\rm Ga}$-3O$_{\rm As}$)$^{-}$&$3/2$&Er$_{\rm Ga}^{0}$ ($S=3/2$) + O$_{\rm As}^{+}$ ($S=0$) + 2O$_{\rm As}^{-}$ ($S=0$)&4.46\\
(Er$_{\rm Ga}$-3O$_{\rm As}$)$^{2-}$&2&Er$_{\rm Ga}^{0}$ ($S=3/2$) + O$_{\rm As}^{0}$ ($S=1/2$) + 2O$_{\rm As}^{-}$ ($S=0$)&4.03\\
(Er$_{\rm Ga}$-4O$_{\rm As}$)$^{2+}$&$3/2$&Er$_{\rm Ga}^{0}$ ($S=3/2$) + 3O$_{\rm As}^{+}$ ($S=0$) + O$_{\rm As}^{-}$ ($S=0$)&4.85\\
(Er$_{\rm Ga}$-4O$_{\rm As}$)$^{+}$&2&Er$_{\rm Ga}^{0}$ ($S=3/2$) + 2O$_{\rm As}^{+}$ ($S=0$) + O$_{\rm As}^{0}$ ($S=1/2$) + O$_{\rm As}^{-}$ ($S=0$)&5.09\\
(Er$_{\rm Ga}$-4O$_{\rm As}$)$^{0}$&$3/2$&Er$_{\rm Ga}^{0}$ ($S=3/2$) + 2O$_{\rm As}^{+}$ ($S=0$) + 2O$_{\rm As}^{-}$ ($S=0$)&5.58\\
(Er$_{\rm Ga}$-4O$_{\rm As}$)$^{-}$&2&Er$_{\rm Ga}^{0}$ ($S=3/2$) + O$_{\rm As}^{+}$ ($S=0$) + O$_{\rm As}^{0}$ ($S=1/2$) + 2O$_{\rm As}^{-}$ ($S=0$)&5.37\\
(Er$_{\rm Ga}$-4O$_{\rm As}$)$^{2-}$&$3/2$&Er$_{\rm Ga}^{0}$ ($S=3/2$) + O$_{\rm As}^{+}$ ($S=0$) + 3O$_{\rm As}^{-}$ ($S=0$)&5.09\\
(Er$_{i,{\rm Ga}}$-O$_{\rm As}$)$^{2+}$&$3/2$&Er$_{i,{\rm Ga}}^{3+}$ ($S=3/2$) + O$_{\rm As}^{-}$ ($S=0$)&1.52 \\
(Er$_{i,{\rm Ga}}$-O$_{\rm As}$)$^{+}$&2&Er$_{i,{\rm Ga}}^{2+}$ ($S=2$) + O$_{\rm As}^{-}$ ($S=0$)&2.02\\
(Er$_{i,{\rm Ga}}$-O$_{\rm As}$)$^0$&$3/2$&Er$_{i,{\rm Ga}}^{+}$ ($S=3/2$) + O$_{\rm As}^{-}$ ($S=0$)&1.68\\
(Er$_{i,{\rm As}}$-O$_{\rm As}$)$^{3+}$&2&Er$_{i,{\rm As}}^{3+}$ ($S=3/2$) + O$_{\rm As}^{0}$ ($S=1/2$)&1.31\\
(Er$_{i,{\rm As}}$-O$_{\rm As}$)$^{2+}$&$3/2$&Er$_{i,{\rm As}}^{3+}$ ($S=3/2$) + O$_{\rm As}^{-}$ ($S=0$)&2.08\\
(Er$_{i,{\rm As}}$-O$_{\rm As}$)$^+$&2&Er$_{i,{\rm As}}^{2+}$ ($S=2$) + O$_{\rm As}^{-}$ ($S=0$)&2.58\\
(Er$_{i,{\rm As}}$-O$_{\rm As}$)$^0$&$3/2$&Er$_{i,{\rm As}}^{+}$ ($S=3/2$) + O$_{\rm As}^{-}$ ($S=0$)&2.91
\end{tabular}
\end{ruledtabular}
\begin{flushleft}
 $^a$The spin of the constituent Ga$_{\rm As}$, $V_{\rm Ga}$, or $V_{\rm As}$ in a given charge state can be different from that when it is in isolation (reported in Ref.~\citenum{Hoang2026JPCM}), depending on the local lattice environment of the constituent in the defect complex.
\end{flushleft}
\end{table*}

\renewcommand\thetable{S2}
\begin{table}
\caption{Defect energy levels [$\epsilon(q/q')$, in eV, with respect to the VBM] induced by Er-related defects.}\label{tab;defectlevel}
\begin{center}
\begin{ruledtabular}
\begin{tabular}{ll}
Defect & Defect energy levels$^{a}$ \\
\colrule
Er$_{\rm As}$ & $(2+/+) = 0.27$, $(+/0) = 0.42$ \\
Er$_{i,{\rm Ga}}$ & $(3+/2+) = 0.74$, $(2+/+) = 0.90$ \\
Er$_{\rm Ga}$-Ga$_{\rm As}$ & $(+/0) = 0.03$, $(0/-) = 1.33$  \\
Er$_{\rm Ga}$-As$_{\rm Ga}$ & $(2+/+) = 0.42$, $(+/0) = 0.79$  \\
Er$_{\rm Ga}$-$V_{\rm Ga}$ & $(+/0) = 0.06$, $(0/-) = 0.34$, $(-/2-) = 0.96$, \\
&$(2-/3-) = 1.25$ \\
Er$_{\rm Ga}$-$V_{\rm As}$ & $(+/0) = 0.62$, $(0/-) = 0.80$, $(-/2-) = 1.20$ \\
Er$_{\rm Ga}$-O$_{i,{\rm tet}}$ & $(+/0) = 0.04$, $(0/-) = 0.20$, $(-/2-) = 0.50$ \\
Er$_{\rm Ga}$-O$_{\rm As}$-O$_{i,{\rm brid}}$ & $(+/0) = 1.06$, $(0/-) = 1.31$  \\
Er$_{i,{\rm Ga}}$-O$_{\rm As}$ & $(2+/+) = 0.24$, $(+/0) = 1.23$ \\
Er$_{i,{\rm As}}$-O$_{\rm As}$ & $(3+/2+) = 0.22$, $(2+/+) = 0.52$, \\
& $(+/0) = 0.89$ \\
Er$_{\rm Ga}$-O$_{\rm As}$ & $(+/0) = 0.80$, $(0/-) = 0.85$  \\
Er$_{\rm Ga}$-2O$_{\rm As}$ & $(2+/0) = 0.92$, $(0/-) = 1.24$, \\
& $(-/2-) = 1.42$; $(2+/+) = 0.93$, \\
& $(+/0) = 0.90$\\
Er$_{\rm Ga}$-3O$_{\rm As}$ & $(2+/+) = 0.77$, $(+/0) = 0.80$, \\
& $(0/-) = 0.80$, $(-/2-) = 1.42$  \\
Er$_{\rm Ga}$-4O$_{\rm As}$ & $(2+/0) = 0.62$, $(0/-) = 1.20$, \\
& $(-/2-) = 1.27$; $(2+/+) = 0.74$, \\
& $(+/0) = 0.51$ 
\end{tabular}
\end{ruledtabular}
\end{center}
\begin{flushleft}
 $^a$Electronically stable but energetically metastable defect levels, if available, are listed after the semicolon.
\end{flushleft}
\end{table}

\renewcommand\thetable{S3}
\begin{table*}
\caption{Key parameters associated with the transitions considered in this work; see text.}\label{tab;opt}
\begin{center}
\begin{ruledtabular}
\begin{tabular}{lcccccc}
Optical transition & $\Delta E$ (eV) & $\Delta Q$ (amu$^{1/2}${\AA}) & $\hbar \Omega_{\rm e}$ (meV) & $\hbar \Omega_{\rm g}$ (meV) & $g_g$ &  $\tilde{W}_{\rm eg}$ (eV/amu$^{1/2}${\AA})\\
\hline
Er$_{i,\rm Ga}^{2+}$ + $e^-$ $\rightarrow$ Er$_{i,\rm Ga}^{+}$ & 0.62 & 5.47 & 9.26 & 9.69 & 1 & 1.83$\times$10$^{-2}$  \\
(Er$_{\rm Ga}$-Ga$_{\rm As}$)$^0$ + $e^-$ $\rightarrow$ (Er$_{\rm Ga}$-Ga$_{\rm As}$)$^-$ & 0.19 & 2.86 & 11.92 & 10.36 & 1 & 0.87$\times$10$^{-2}$  \\
(Er$_{\rm Ga}$-As$_{\rm Ga}$)$^+$ + $e^-$ $\rightarrow$ (Er$_{\rm Ga}$-As$_{\rm Ga}$)$^0$ & 0.73 & 1.71 & 17.55 & 16.93 & 1 & 2.55$\times$10$^{-2}$ \\
(Er$_{\rm Ga}$-O$_{\rm As}$)$^+$ + $e^-$ $\rightarrow$ (Er$_{\rm Ga}$-O$_{\rm As}$)$^0$ & 0.72 & 6.05 & 12.24 & 13.34 & 1 & 3.34$\times$10$^{-2}$ \\
(Er$_{\rm Ga}$-O$_{\rm As}$)$^0$ + $e^-$ $\rightarrow$ (Er$_{\rm Ga}$-O$_{\rm As}$)$^-$ & 0.67 & 3.72 & 15.03 & 17.25 & 1 & 6.36$\times$10$^{-2}$ \\
(Er$_{\rm Ga}$-2O$_{\rm As}$)$^{2+}$ + $e^-$ $\rightarrow$ (Er$_{\rm Ga}$-2O$_{\rm As}$)$^+$ & 0.58 & 4.62 & 14.65 & 13.67 & 2 & 6.50$\times$10$^{-2}$ \\
(Er$_{\rm Ga}$-2O$_{\rm As}$)$^+$ + $e^-$ $\rightarrow$ (Er$_{\rm Ga}$-2O$_{\rm As}$)$^0$ & 0.61 & 6.05 & 10.30 & 11.85 & 1 & 3.94$\times$10$^{-2}$ \\
(Er$_{\rm Ga}$-2O$_{\rm As}$)$^0$ + $e^-$ $\rightarrow$ (Er$_{\rm Ga}$-2O$_{\rm As}$)$^-$ & 0.27 & 4.76 & 13.78 & 13.05 & 1 & 6.71$\times$10$^{-2}$ \\
(Er$_{\rm Ga}$-3O$_{\rm As}$)$^+$ + $e^-$ $\rightarrow$ (Er$_{\rm Ga}$-3O$_{\rm As}$)$^0$ & 0.72 & 3.34 & 14.88 & 17.52 & 2 & 2.36$\times$10$^{-2}$ \\
(Er$_{\rm Ga}$-3O$_{\rm As}$)$^0$ + $e^-$ $\rightarrow$ (Er$_{\rm Ga}$-3O$_{\rm As}$)$^-$ & 0.71 & 9.30 & 11.37 & 10.81 & 1 & 1.96$\times$10$^{-2}$ \\
(Er$_{\rm Ga}$-4O$_{\rm As}$)$^0$ + $e^-$ $\rightarrow$ (Er$_{\rm Ga}$-4O$_{\rm As}$)$^-$ & 0.32 & 4.45 & 14.57 & 12.48 & 2 & 8.60$\times$10$^{-2}$ \\
\end{tabular}
\end{ruledtabular}
\end{center}
\end{table*}

\renewcommand{\thefigure}{S3}
\begin{figure}
\vspace{0.2cm}
\includegraphics*[width=\linewidth]{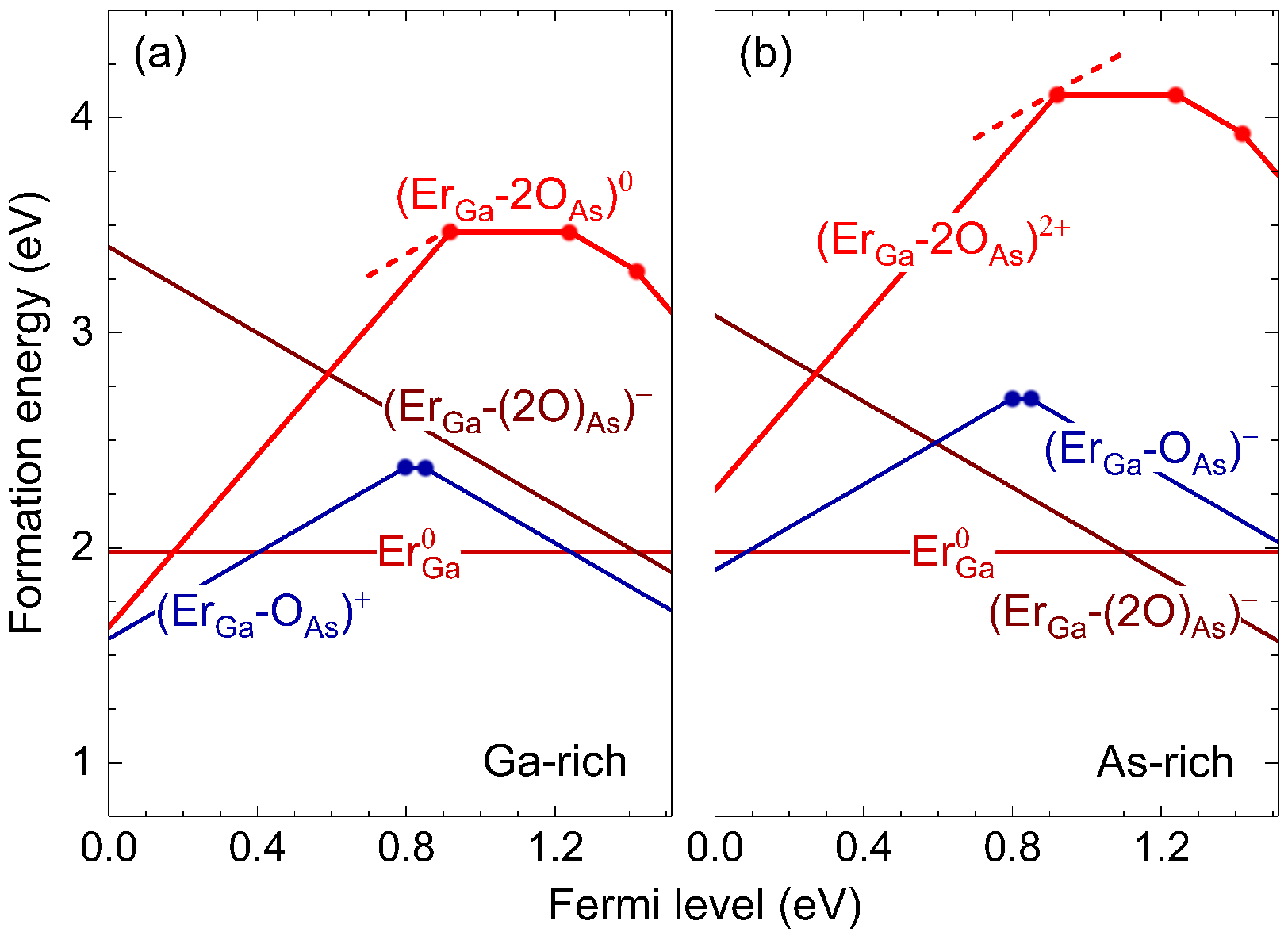}
\caption{Er$_{\rm Ga}$-2O$_{\rm As}$ and the lowest-energy Er-related defects under either the extreme Ga-rich or As-rich condition.}
\label{fig;compared} 
\end{figure}

\renewcommand{\thefigure}{S4}
\begin{figure}
\vspace{0.2cm}
\includegraphics*[width=\linewidth]{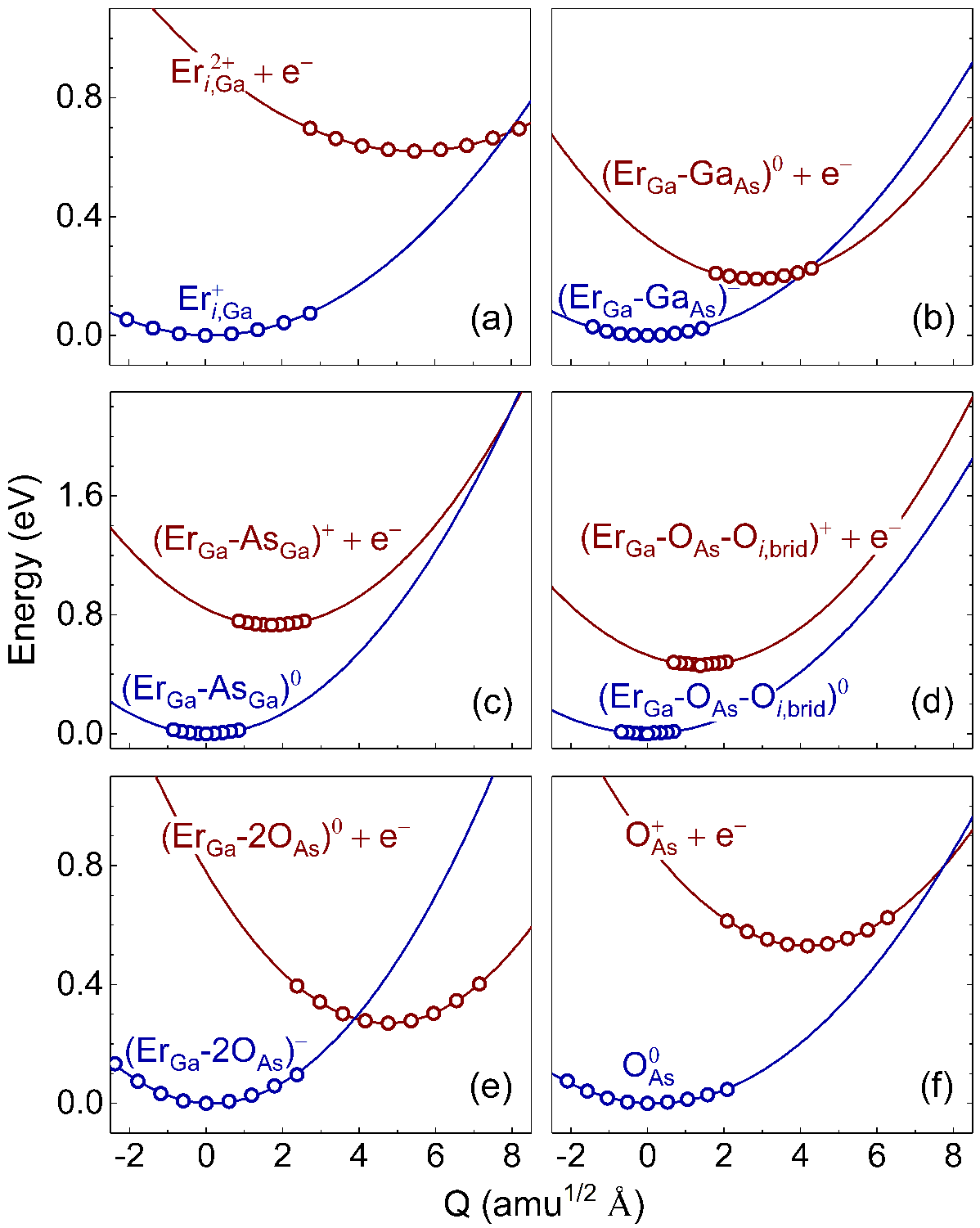} 
\caption{Configuration coordinate diagrams for transitions involving select Er-related defect complexes in GaAs.}
\label{fig;morecc} 
\end{figure}

\renewcommand{\thefigure}{S5}
\begin{figure}
\vspace{0.2cm}
\includegraphics*[width=\linewidth]{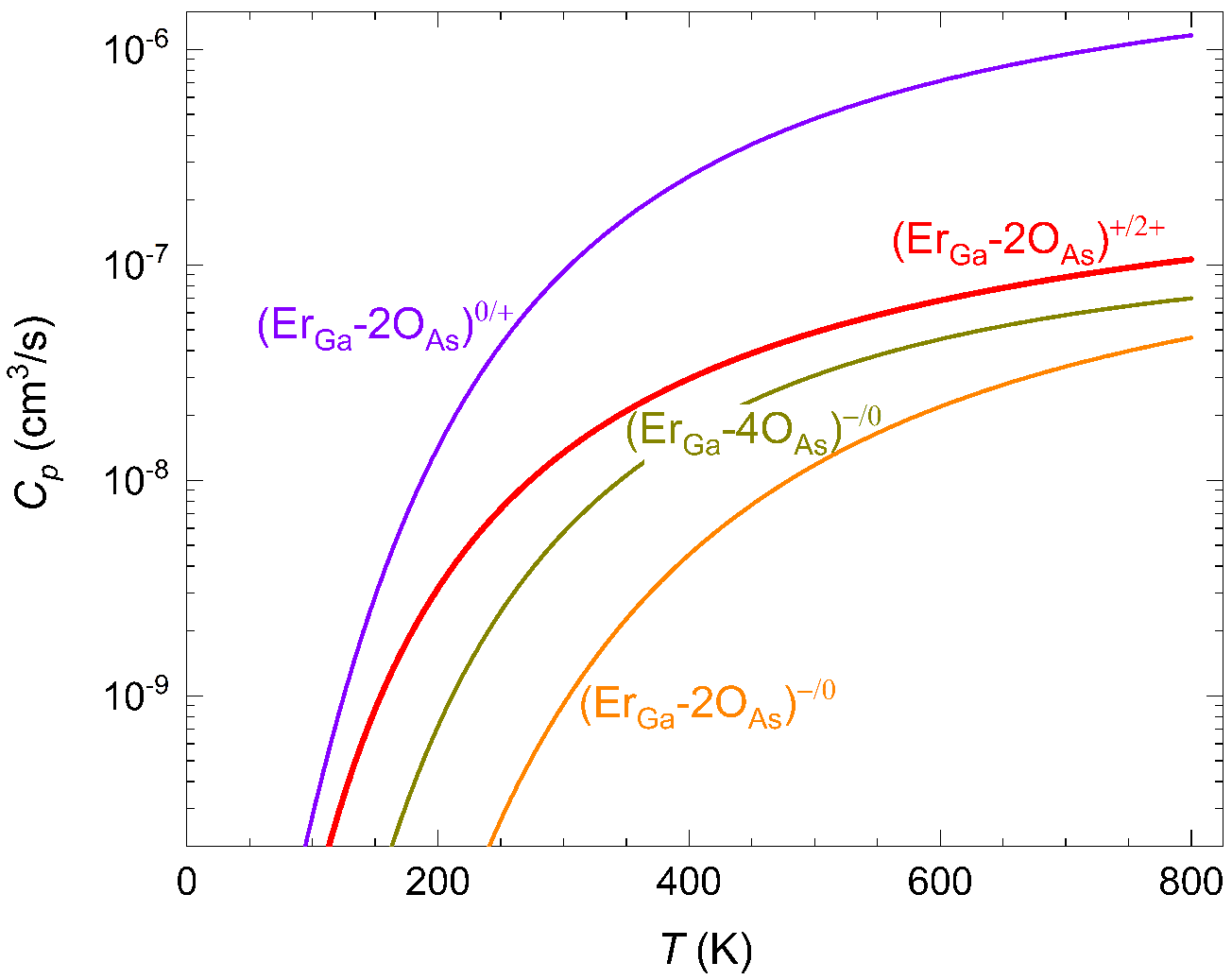} 
\caption{Nonradiative hole capture coefficient ($C_p$) at select Er-related defect centers in GaAs; see text.}
\label{fig;Cp} 
\end{figure}


\begin{thebibliography}{71}%
\makeatletter
\providecommand \@ifxundefined [1]{%
 \@ifx{#1\undefined}
}%
\providecommand \@ifnum [1]{%
 \ifnum #1\expandafter \@firstoftwo
 \else \expandafter \@secondoftwo
 \fi
}%
\providecommand \@ifx [1]{%
 \ifx #1\expandafter \@firstoftwo
 \else \expandafter \@secondoftwo
 \fi
}%
\providecommand \natexlab [1]{#1}%
\providecommand \enquote  [1]{``#1''}%
\providecommand \bibnamefont  [1]{#1}%
\providecommand \bibfnamefont [1]{#1}%
\providecommand \citenamefont [1]{#1}%
\providecommand \href@noop [0]{\@secondoftwo}%
\providecommand \href [0]{\begingroup \@sanitize@url \@href}%
\providecommand \@href[1]{\@@startlink{#1}\@@href}%
\providecommand \@@href[1]{\endgroup#1\@@endlink}%
\providecommand \@sanitize@url [0]{\catcode `\\12\catcode `\$12\catcode
  `\&12\catcode `\#12\catcode `\^12\catcode `\_12\catcode `\%12\relax}%
\providecommand \@@startlink[1]{}%
\providecommand \@@endlink[0]{}%
\providecommand \url  [0]{\begingroup\@sanitize@url \@url }%
\providecommand \@url [1]{\endgroup\@href {#1}{\urlprefix }}%
\providecommand \urlprefix  [0]{URL }%
\providecommand \Eprint [0]{\href }%
\providecommand \doibase [0]{https://doi.org/}%
\providecommand \selectlanguage [0]{\@gobble}%
\providecommand \bibinfo  [0]{\@secondoftwo}%
\providecommand \bibfield  [0]{\@secondoftwo}%
\providecommand \translation [1]{[#1]}%
\providecommand \BibitemOpen [0]{}%
\providecommand \bibitemStop [0]{}%
\providecommand \bibitemNoStop [0]{.\EOS\space}%
\providecommand \EOS [0]{\spacefactor3000\relax}%
\providecommand \BibitemShut  [1]{\csname bibitem#1\endcsname}%
\let\auto@bib@innerbib\@empty
\bibitem [{\citenamefont {O'Donnell}\ and\ \citenamefont
  {Dierolf}(2010)}]{ODonnell2010Book}%
  \BibitemOpen
  \bibinfo {editor} {\bibfnamefont {K.}~\bibnamefont {O'Donnell}}\ and\
  \bibinfo {editor} {\bibfnamefont {V.}~\bibnamefont {Dierolf}},\ eds.,\
  \href@noop {} {\emph {\bibinfo {title} {{Rare Earth Doped III-Nitrides for
  Optoelectronic and Spintronic Applications}}}},\ \bibinfo {series} {Topics in
  Applied Physics}, Vol.\ \bibinfo {volume} {124}\ (\bibinfo  {publisher}
  {Springer},\ \bibinfo {address} {Dordrecht},\ \bibinfo {year}
  {2010})\BibitemShut {NoStop}%
\bibitem [{\citenamefont {Yin}\ \emph {et~al.}(2013)\citenamefont {Yin},
  \citenamefont {Rancic}, \citenamefont {{de Boo}}, \citenamefont {Stavrias},
  \citenamefont {McCallum}, \citenamefont {Sellars},\ and\ \citenamefont
  {Rogge}}]{Yin2013Nature}%
  \BibitemOpen
  \bibfield  {author} {\bibinfo {author} {\bibfnamefont {C.}~\bibnamefont
  {Yin}}, \bibinfo {author} {\bibfnamefont {M.}~\bibnamefont {Rancic}},
  \bibinfo {author} {\bibfnamefont {G.~G.}\ \bibnamefont {{de Boo}}}, \bibinfo
  {author} {\bibfnamefont {N.}~\bibnamefont {Stavrias}}, \bibinfo {author}
  {\bibfnamefont {J.~C.}\ \bibnamefont {McCallum}}, \bibinfo {author}
  {\bibfnamefont {M.~J.}\ \bibnamefont {Sellars}},\ and\ \bibinfo {author}
  {\bibfnamefont {S.}~\bibnamefont {Rogge}},\ }\bibfield  {title} {\bibinfo
  {title} {Optical addressing of an individual erbium ion in silicon},\ }\href
  {https://doi.org/10.1038/nature12081} {\bibfield  {journal} {\bibinfo
  {journal} {Nature}\ }\textbf {\bibinfo {volume} {497}},\ \bibinfo {pages}
  {91} (\bibinfo {year} {2013})}\BibitemShut {NoStop}%
\bibitem [{\citenamefont {Ogawa}\ \emph {et~al.}(2020)\citenamefont {Ogawa},
  \citenamefont {Tatebayashi}, \citenamefont {Fujioka}, \citenamefont
  {Higashi}, \citenamefont {Fujita}, \citenamefont {Noda}, \citenamefont
  {Timmerman}, \citenamefont {Ichikawa},\ and\ \citenamefont
  {Fujiwara}}]{Ogawa2020APL}%
  \BibitemOpen
  \bibfield  {author} {\bibinfo {author} {\bibfnamefont {M.}~\bibnamefont
  {Ogawa}}, \bibinfo {author} {\bibfnamefont {J.}~\bibnamefont {Tatebayashi}},
  \bibinfo {author} {\bibfnamefont {N.}~\bibnamefont {Fujioka}}, \bibinfo
  {author} {\bibfnamefont {R.}~\bibnamefont {Higashi}}, \bibinfo {author}
  {\bibfnamefont {M.}~\bibnamefont {Fujita}}, \bibinfo {author} {\bibfnamefont
  {S.}~\bibnamefont {Noda}}, \bibinfo {author} {\bibfnamefont {D.}~\bibnamefont
  {Timmerman}}, \bibinfo {author} {\bibfnamefont {S.}~\bibnamefont
  {Ichikawa}},\ and\ \bibinfo {author} {\bibfnamefont {Y.}~\bibnamefont
  {Fujiwara}},\ }\bibfield  {title} {\bibinfo {title} {Quantitative evaluation
  of enhanced {Er} luminescence in {GaAs}-based two-dimensional photonic
  crystal nanocavities},\ }\href {https://doi.org/10.1063/1.5142778} {\bibfield
   {journal} {\bibinfo  {journal} {Appl. Phys. Lett.}\ }\textbf {\bibinfo
  {volume} {116}},\ \bibinfo {pages} {181102} (\bibinfo {year}
  {2020})}\BibitemShut {NoStop}%
\bibitem [{\citenamefont {Higashi}\ \emph {et~al.}(2020)\citenamefont
  {Higashi}, \citenamefont {Ogawa}, \citenamefont {Tatebayashi}, \citenamefont
  {Fujioka}, \citenamefont {Timmerman}, \citenamefont {Ichikawa},\ and\
  \citenamefont {Fujiwara}}]{Higashi2020JAP}%
  \BibitemOpen
  \bibfield  {author} {\bibinfo {author} {\bibfnamefont {R.}~\bibnamefont
  {Higashi}}, \bibinfo {author} {\bibfnamefont {M.}~\bibnamefont {Ogawa}},
  \bibinfo {author} {\bibfnamefont {J.}~\bibnamefont {Tatebayashi}}, \bibinfo
  {author} {\bibfnamefont {N.}~\bibnamefont {Fujioka}}, \bibinfo {author}
  {\bibfnamefont {D.}~\bibnamefont {Timmerman}}, \bibinfo {author}
  {\bibfnamefont {S.}~\bibnamefont {Ichikawa}},\ and\ \bibinfo {author}
  {\bibfnamefont {Y.}~\bibnamefont {Fujiwara}},\ }\bibfield  {title} {\bibinfo
  {title} {Enhancement of {Er} luminescence in microdisk resonators made of
  {Er,O-codoped GaAs}},\ }\href {https://doi.org/10.1063/1.5144159} {\bibfield
  {journal} {\bibinfo  {journal} {J. Appl. Phys.}\ }\textbf {\bibinfo {volume}
  {127}},\ \bibinfo {pages} {233101} (\bibinfo {year} {2020})}\BibitemShut
  {NoStop}%
\bibitem [{\citenamefont {Fang}\ \emph {et~al.}(2023)\citenamefont {Fang},
  \citenamefont {Tatebayashi}, \citenamefont {Homi}, \citenamefont {Ogawa},
  \citenamefont {Kajii}, \citenamefont {Kondow}, \citenamefont {Kitamura},
  \citenamefont {Mitchell}, \citenamefont {Ichikawa},\ and\ \citenamefont
  {Fujiwara}}]{Fang2023OC}%
  \BibitemOpen
  \bibfield  {author} {\bibinfo {author} {\bibfnamefont {Z.}~\bibnamefont
  {Fang}}, \bibinfo {author} {\bibfnamefont {J.}~\bibnamefont {Tatebayashi}},
  \bibinfo {author} {\bibfnamefont {R.}~\bibnamefont {Homi}}, \bibinfo {author}
  {\bibfnamefont {M.}~\bibnamefont {Ogawa}}, \bibinfo {author} {\bibfnamefont
  {H.}~\bibnamefont {Kajii}}, \bibinfo {author} {\bibfnamefont
  {M.}~\bibnamefont {Kondow}}, \bibinfo {author} {\bibfnamefont
  {K.}~\bibnamefont {Kitamura}}, \bibinfo {author} {\bibfnamefont
  {B.}~\bibnamefont {Mitchell}}, \bibinfo {author} {\bibfnamefont
  {S.}~\bibnamefont {Ichikawa}},\ and\ \bibinfo {author} {\bibfnamefont
  {Y.}~\bibnamefont {Fujiwara}},\ }\bibfield  {title} {\bibinfo {title}
  {Enhancement of {Er} luminescence from bridge-type photonic crystal
  nanocavities with {Er, O-co-doped GaAs}},\ }\href
  {https://doi.org/10.1364/OPTCON.501666} {\bibfield  {journal} {\bibinfo
  {journal} {Opt. Continuum}\ }\textbf {\bibinfo {volume} {2}},\ \bibinfo
  {pages} {2178} (\bibinfo {year} {2023})}\BibitemShut {NoStop}%
\bibitem [{\citenamefont {Fang}\ \emph {et~al.}(2025)\citenamefont {Fang},
  \citenamefont {Kajii}, \citenamefont {Kondow}, \citenamefont {Fujiwara},\
  and\ \citenamefont {Tatebayashi}}]{Fang2025JJAP}%
  \BibitemOpen
  \bibfield  {author} {\bibinfo {author} {\bibfnamefont {Z.}~\bibnamefont
  {Fang}}, \bibinfo {author} {\bibfnamefont {H.}~\bibnamefont {Kajii}},
  \bibinfo {author} {\bibfnamefont {M.}~\bibnamefont {Kondow}}, \bibinfo
  {author} {\bibfnamefont {Y.}~\bibnamefont {Fujiwara}},\ and\ \bibinfo
  {author} {\bibfnamefont {J.}~\bibnamefont {Tatebayashi}},\ }\bibfield
  {title} {\bibinfo {title} {Demonstration of enhanced {Er} luminescence in
  nanobeam photonic crystal nanocavities based on {Er,O-codoped GaAs}},\ }\href
  {https://doi.org/10.35848/1347-4065/adaff3} {\bibfield  {journal} {\bibinfo
  {journal} {Jpn. J. Appl. Phys.}\ }\textbf {\bibinfo {volume} {64}},\ \bibinfo
  {pages} {025001} (\bibinfo {year} {2025})}\BibitemShut {NoStop}%
\bibitem [{\citenamefont {Thiel}\ \emph {et~al.}(2011)\citenamefont {Thiel},
  \citenamefont {B\"{o}ttger},\ and\ \citenamefont {Cone}}]{Thiel2011JL}%
  \BibitemOpen
  \bibfield  {author} {\bibinfo {author} {\bibfnamefont {C.}~\bibnamefont
  {Thiel}}, \bibinfo {author} {\bibfnamefont {T.}~\bibnamefont {B\"{o}ttger}},\
  and\ \bibinfo {author} {\bibfnamefont {R.}~\bibnamefont {Cone}},\ }\bibfield
  {title} {\bibinfo {title} {Rare-earth-doped materials for applications in
  quantum information storage and signal processing},\ }\href
  {https://doi.org/10.1016/j.jlumin.2010.12.015} {\bibfield  {journal}
  {\bibinfo  {journal} {J. Lumin.}\ }\textbf {\bibinfo {volume} {131}},\
  \bibinfo {pages} {353} (\bibinfo {year} {2011})}\BibitemShut {NoStop}%
\bibitem [{\citenamefont {Kunkel}\ and\ \citenamefont
  {Goldner}(2018)}]{Kunkel2018ZAAC}%
  \BibitemOpen
  \bibfield  {author} {\bibinfo {author} {\bibfnamefont {N.}~\bibnamefont
  {Kunkel}}\ and\ \bibinfo {author} {\bibfnamefont {P.}~\bibnamefont
  {Goldner}},\ }\bibfield  {title} {\bibinfo {title} {{Recent Advances in Rare
  Earth Doped Inorganic Crystalline Materials for Quantum Information
  Processing}},\ }\href {https://doi.org/10.1002/zaac.201700425} {\bibfield
  {journal} {\bibinfo  {journal} {Z. Anorg. Allg. Chem.}\ }\textbf {\bibinfo
  {volume} {644}},\ \bibinfo {pages} {66} (\bibinfo {year} {2018})}\BibitemShut
  {NoStop}%
\bibitem [{\citenamefont {Ennen}\ \emph {et~al.}(1983)\citenamefont {Ennen},
  \citenamefont {Schneider}, \citenamefont {Pomrenke},\ and\ \citenamefont
  {Axmann}}]{Ennen1983APL}%
  \BibitemOpen
  \bibfield  {author} {\bibinfo {author} {\bibfnamefont {H.}~\bibnamefont
  {Ennen}}, \bibinfo {author} {\bibfnamefont {J.}~\bibnamefont {Schneider}},
  \bibinfo {author} {\bibfnamefont {G.}~\bibnamefont {Pomrenke}},\ and\
  \bibinfo {author} {\bibfnamefont {A.}~\bibnamefont {Axmann}},\ }\bibfield
  {title} {\bibinfo {title} {{1.54‐$\mu$m luminescence of erbium‐implanted
  III‐V semiconductors and silicon}},\ }\href
  {https://doi.org/10.1063/1.94190} {\bibfield  {journal} {\bibinfo  {journal}
  {Appl. Phys. Lett.}\ }\textbf {\bibinfo {volume} {43}},\ \bibinfo {pages}
  {943} (\bibinfo {year} {1983})}\BibitemShut {NoStop}%
\bibitem [{\citenamefont {Pomrenke}\ \emph {et~al.}(1986)\citenamefont
  {Pomrenke}, \citenamefont {Ennen},\ and\ \citenamefont
  {Haydl}}]{Pomrenke1986JAP}%
  \BibitemOpen
  \bibfield  {author} {\bibinfo {author} {\bibfnamefont {G.~S.}\ \bibnamefont
  {Pomrenke}}, \bibinfo {author} {\bibfnamefont {H.}~\bibnamefont {Ennen}},\
  and\ \bibinfo {author} {\bibfnamefont {W.}~\bibnamefont {Haydl}},\ }\bibfield
   {title} {\bibinfo {title} {Photoluminescence optimization and
  characteristics of the rare‐earth element erbium implanted in {GaAs, InP,
  and GaP}},\ }\href {https://doi.org/10.1063/1.336619} {\bibfield  {journal}
  {\bibinfo  {journal} {J. Appl. Phys.}\ }\textbf {\bibinfo {volume} {59}},\
  \bibinfo {pages} {601} (\bibinfo {year} {1986})}\BibitemShut {NoStop}%
\bibitem [{\citenamefont {Smith}\ \emph {et~al.}(1987)\citenamefont {Smith},
  \citenamefont {Müller}, \citenamefont {Ennen}, \citenamefont {Wennekers},\
  and\ \citenamefont {Maier}}]{Smith1987APL}%
  \BibitemOpen
  \bibfield  {author} {\bibinfo {author} {\bibfnamefont {R.~S.}\ \bibnamefont
  {Smith}}, \bibinfo {author} {\bibfnamefont {H.~D.}\ \bibnamefont {Müller}},
  \bibinfo {author} {\bibfnamefont {H.}~\bibnamefont {Ennen}}, \bibinfo
  {author} {\bibfnamefont {P.}~\bibnamefont {Wennekers}},\ and\ \bibinfo
  {author} {\bibfnamefont {M.}~\bibnamefont {Maier}},\ }\bibfield  {title}
  {\bibinfo {title} {{Erbium doping of molecular beam epitaxial GaAs}},\ }\href
  {https://doi.org/10.1063/1.98127} {\bibfield  {journal} {\bibinfo  {journal}
  {Appl. Phys. Lett.}\ }\textbf {\bibinfo {volume} {50}},\ \bibinfo {pages}
  {49} (\bibinfo {year} {1987})}\BibitemShut {NoStop}%
\bibitem [{\citenamefont {Ennen}\ \emph {et~al.}(1987)\citenamefont {Ennen},
  \citenamefont {Wagner}, \citenamefont {Müller},\ and\ \citenamefont
  {Smith}}]{Ennen1987JAP}%
  \BibitemOpen
  \bibfield  {author} {\bibinfo {author} {\bibfnamefont {H.}~\bibnamefont
  {Ennen}}, \bibinfo {author} {\bibfnamefont {J.}~\bibnamefont {Wagner}},
  \bibinfo {author} {\bibfnamefont {H.~D.}\ \bibnamefont {Müller}},\ and\
  \bibinfo {author} {\bibfnamefont {R.~S.}\ \bibnamefont {Smith}},\ }\bibfield
  {title} {\bibinfo {title} {Photoluminescence excitation measurements on
  {GaAs:Er} grown by molecular‐beam epitaxy},\ }\href
  {https://doi.org/10.1063/1.338353} {\bibfield  {journal} {\bibinfo  {journal}
  {J. Appl. Phys.}\ }\textbf {\bibinfo {volume} {61}},\ \bibinfo {pages} {4877}
  (\bibinfo {year} {1987})}\BibitemShut {NoStop}%
\bibitem [{\citenamefont {Thonke}\ \emph {et~al.}(1988)\citenamefont {Thonke},
  \citenamefont {Hermann},\ and\ \citenamefont {Schneider}}]{Thonke1988JPC}%
  \BibitemOpen
  \bibfield  {author} {\bibinfo {author} {\bibfnamefont {K.}~\bibnamefont
  {Thonke}}, \bibinfo {author} {\bibfnamefont {H.~U.}\ \bibnamefont
  {Hermann}},\ and\ \bibinfo {author} {\bibfnamefont {J.}~\bibnamefont
  {Schneider}},\ }\bibfield  {title} {\bibinfo {title} {{A Zeeman study of the
  1.54{$\mu$m} transition in molecular beam epitaxial GaAs:Er}},\ }\href
  {https://doi.org/10.1088/0022-3719/21/34/020} {\bibfield  {journal} {\bibinfo
   {journal} {J. Phys. C: Solid State Phys.}\ }\textbf {\bibinfo {volume}
  {21}},\ \bibinfo {pages} {5881} (\bibinfo {year} {1988})}\BibitemShut
  {NoStop}%
\bibitem [{\citenamefont {Takahei}\ \emph {et~al.}(1989)\citenamefont
  {Takahei}, \citenamefont {Whitney}, \citenamefont {Nakagome},\ and\
  \citenamefont {Uwai}}]{Takahei1989JAP}%
  \BibitemOpen
  \bibfield  {author} {\bibinfo {author} {\bibfnamefont {K.}~\bibnamefont
  {Takahei}}, \bibinfo {author} {\bibfnamefont {P.~S.}\ \bibnamefont
  {Whitney}}, \bibinfo {author} {\bibfnamefont {H.}~\bibnamefont {Nakagome}},\
  and\ \bibinfo {author} {\bibfnamefont {K.}~\bibnamefont {Uwai}},\ }\bibfield
  {title} {\bibinfo {title} {{Temperature dependence of intra‐$4f$‐shell
  photo‐ and electroluminescence spectra for erbium‐doped GaAs}},\ }\href
  {https://doi.org/10.1063/1.343019} {\bibfield  {journal} {\bibinfo  {journal}
  {J. Appl. Phys.}\ }\textbf {\bibinfo {volume} {65}},\ \bibinfo {pages} {1257}
  (\bibinfo {year} {1989})}\BibitemShut {NoStop}%
\bibitem [{\citenamefont {Kozanecki}\ \emph {et~al.}(1991)\citenamefont
  {Kozanecki}, \citenamefont {Chan}, \citenamefont {Jeynes}, \citenamefont
  {Sealy},\ and\ \citenamefont {Homewood}}]{Kozanecki1991SSC}%
  \BibitemOpen
  \bibfield  {author} {\bibinfo {author} {\bibfnamefont {A.}~\bibnamefont
  {Kozanecki}}, \bibinfo {author} {\bibfnamefont {M.}~\bibnamefont {Chan}},
  \bibinfo {author} {\bibfnamefont {C.}~\bibnamefont {Jeynes}}, \bibinfo
  {author} {\bibfnamefont {B.}~\bibnamefont {Sealy}},\ and\ \bibinfo {author}
  {\bibfnamefont {K.}~\bibnamefont {Homewood}},\ }\bibfield  {title} {\bibinfo
  {title} {Lattice location of erbium implanted into {GaAs}},\ }\href
  {https://doi.org/10.1016/0038-1098(91)90860-X} {\bibfield  {journal}
  {\bibinfo  {journal} {Solid State Commun.}\ }\textbf {\bibinfo {volume}
  {78}},\ \bibinfo {pages} {763} (\bibinfo {year} {1991})}\BibitemShut
  {NoStop}%
\bibitem [{\citenamefont {Nakata}\ \emph {et~al.}(1992)\citenamefont {Nakata},
  \citenamefont {Taniguchi},\ and\ \citenamefont {Takahei}}]{Nakata1992APL}%
  \BibitemOpen
  \bibfield  {author} {\bibinfo {author} {\bibfnamefont {J.}~\bibnamefont
  {Nakata}}, \bibinfo {author} {\bibfnamefont {M.}~\bibnamefont {Taniguchi}},\
  and\ \bibinfo {author} {\bibfnamefont {K.}~\bibnamefont {Takahei}},\
  }\bibfield  {title} {\bibinfo {title} {{Direct evidence of Er atoms occupying
  an interstitial site in metalorganic chemical vapor deposition‐grown
  GaAs:Er}},\ }\href {https://doi.org/10.1063/1.108102} {\bibfield  {journal}
  {\bibinfo  {journal} {Appl. Phys. Lett.}\ }\textbf {\bibinfo {volume} {61}},\
  \bibinfo {pages} {2665} (\bibinfo {year} {1992})}\BibitemShut {NoStop}%
\bibitem [{\citenamefont {Nakata}\ \emph {et~al.}(1995)\citenamefont {Nakata},
  \citenamefont {Jourdan}, \citenamefont {Yamaguchi}, \citenamefont {Takahei},
  \citenamefont {Yamamoto},\ and\ \citenamefont {Kido}}]{Nakata1995JAP}%
  \BibitemOpen
  \bibfield  {author} {\bibinfo {author} {\bibfnamefont {J.}~\bibnamefont
  {Nakata}}, \bibinfo {author} {\bibfnamefont {N.}~\bibnamefont {Jourdan}},
  \bibinfo {author} {\bibfnamefont {H.}~\bibnamefont {Yamaguchi}}, \bibinfo
  {author} {\bibfnamefont {K.}~\bibnamefont {Takahei}}, \bibinfo {author}
  {\bibfnamefont {Y.}~\bibnamefont {Yamamoto}},\ and\ \bibinfo {author}
  {\bibfnamefont {Y.}~\bibnamefont {Kido}},\ }\bibfield  {title} {\bibinfo
  {title} {{Structural analysis of erbium sheet‐doped GaAs grown by
  molecular‐beam epitaxy, with ion channeling followed by Monte Carlo
  simulation}},\ }\href {https://doi.org/10.1063/1.358660} {\bibfield
  {journal} {\bibinfo  {journal} {J. Appl. Phys.}\ }\textbf {\bibinfo {volume}
  {77}},\ \bibinfo {pages} {3095} (\bibinfo {year} {1995})}\BibitemShut
  {NoStop}%
\bibitem [{\citenamefont {Alves}\ \emph {et~al.}(1998)\citenamefont {Alves},
  \citenamefont {{da Silva}}, \citenamefont {Soares}, \citenamefont {Henry},
  \citenamefont {Gwilliam}, \citenamefont {Sealy}, \citenamefont {Freitag},
  \citenamefont {Vianden},\ and\ \citenamefont {Stievenard}}]{Alves1998NIMPR}%
  \BibitemOpen
  \bibfield  {author} {\bibinfo {author} {\bibfnamefont {E.}~\bibnamefont
  {Alves}}, \bibinfo {author} {\bibfnamefont {M.~F.}\ \bibnamefont {{da
  Silva}}}, \bibinfo {author} {\bibfnamefont {J.~C.}\ \bibnamefont {Soares}},
  \bibinfo {author} {\bibfnamefont {M.~O.}\ \bibnamefont {Henry}}, \bibinfo
  {author} {\bibfnamefont {R.}~\bibnamefont {Gwilliam}}, \bibinfo {author}
  {\bibfnamefont {B.~J.}\ \bibnamefont {Sealy}}, \bibinfo {author}
  {\bibfnamefont {K.}~\bibnamefont {Freitag}}, \bibinfo {author} {\bibfnamefont
  {R.}~\bibnamefont {Vianden}},\ and\ \bibinfo {author} {\bibfnamefont
  {D.}~\bibnamefont {Stievenard}},\ }\bibfield  {title} {\bibinfo {title}
  {Lattice site location of thulium and erbium implanted {GaAs}},\ }\href
  {https://doi.org/10.1016/S0168-583X(97)00715-5} {\bibfield  {journal}
  {\bibinfo  {journal} {Nucl. Instrum. Methods Phys. Res. B}\ }\textbf
  {\bibinfo {volume} {136--138}},\ \bibinfo {pages} {421} (\bibinfo {year}
  {1998})}\BibitemShut {NoStop}%
\bibitem [{\citenamefont {Wahl}\ \emph {et~al.}(1999)\citenamefont {Wahl},
  \citenamefont {Vantomme}, \citenamefont {Langouche},\ and\ \citenamefont
  {{ISOLDE collaboration}}}]{Wahl1999NIMPR}%
  \BibitemOpen
  \bibfield  {author} {\bibinfo {author} {\bibfnamefont {U.}~\bibnamefont
  {Wahl}}, \bibinfo {author} {\bibfnamefont {A.}~\bibnamefont {Vantomme}},
  \bibinfo {author} {\bibfnamefont {G.}~\bibnamefont {Langouche}},\ and\
  \bibinfo {author} {\bibfnamefont {T.}~\bibnamefont {{ISOLDE
  collaboration}}},\ }\bibfield  {title} {\bibinfo {title} {Lattice sites and
  damage annealing of {Er} in low-dose implanted {GaAs}},\ }\href
  {https://doi.org/10.1016/S0168-583X(98)00673-9} {\bibfield  {journal}
  {\bibinfo  {journal} {Nucl. Instrum. Methods Phys. Res. B}\ }\textbf
  {\bibinfo {volume} {148}},\ \bibinfo {pages} {492} (\bibinfo {year}
  {1999})}\BibitemShut {NoStop}%
\bibitem [{\citenamefont {Nakagome}\ \emph {et~al.}(1988)\citenamefont
  {Nakagome}, \citenamefont {Uwai},\ and\ \citenamefont
  {Takahei}}]{Nakagome1988APL}%
  \BibitemOpen
  \bibfield  {author} {\bibinfo {author} {\bibfnamefont {H.}~\bibnamefont
  {Nakagome}}, \bibinfo {author} {\bibfnamefont {K.}~\bibnamefont {Uwai}},\
  and\ \bibinfo {author} {\bibfnamefont {K.}~\bibnamefont {Takahei}},\
  }\bibfield  {title} {\bibinfo {title} {Extremely sharp erbium‐related
  intra‐$4f$‐shell photoluminescence of erbium‐doped {GaAs} grown by
  metalorganic chemical vapor deposition},\ }\href
  {https://doi.org/10.1063/1.99807} {\bibfield  {journal} {\bibinfo  {journal}
  {Appl. Phys. Lett.}\ }\textbf {\bibinfo {volume} {53}},\ \bibinfo {pages}
  {1726} (\bibinfo {year} {1988})}\BibitemShut {NoStop}%
\bibitem [{\citenamefont {Taguchi}\ \emph {et~al.}(1993)\citenamefont
  {Taguchi}, \citenamefont {Takahei},\ and\ \citenamefont
  {Nakata}}]{Taguchi1993MRS}%
  \BibitemOpen
  \bibfield  {author} {\bibinfo {author} {\bibfnamefont {A.}~\bibnamefont
  {Taguchi}}, \bibinfo {author} {\bibfnamefont {K.}~\bibnamefont {Takahei}},\
  and\ \bibinfo {author} {\bibfnamefont {J.}~\bibnamefont {Nakata}},\
  }\bibfield  {title} {\bibinfo {title} {Electronic properties and their
  relations to optical properties in rare earth doped {III-V} semiconductors},\
  }in\ \href@noop {} {\emph {\bibinfo {booktitle} {Rare Earth Doped
  Semiconductors}}},\ \bibinfo {series} {Mat. Res. Soc. Symp. Proc.}, Vol.\
  \bibinfo {volume} {301}\ (\bibinfo  {publisher} {Materials Research
  Society},\ \bibinfo {address} {Pittsburgh, PA},\ \bibinfo {year} {1993})\
  pp.\ \bibinfo {pages} {139--150}\BibitemShut {NoStop}%
\bibitem [{\citenamefont {Takahei}\ and\ \citenamefont
  {Taguchi}(1993)}]{Takahei1993JAP}%
  \BibitemOpen
  \bibfield  {author} {\bibinfo {author} {\bibfnamefont {K.}~\bibnamefont
  {Takahei}}\ and\ \bibinfo {author} {\bibfnamefont {A.}~\bibnamefont
  {Taguchi}},\ }\bibfield  {title} {\bibinfo {title} {Selective formation of an
  efficient {Er‐O} luminescence center in {GaAs} by metalorganic chemical
  vapor deposition under an atmosphere containing oxygen},\ }\href
  {https://doi.org/10.1063/1.354757} {\bibfield  {journal} {\bibinfo  {journal}
  {J. Appl. Phys.}\ }\textbf {\bibinfo {volume} {74}},\ \bibinfo {pages} {1979}
  (\bibinfo {year} {1993})}\BibitemShut {NoStop}%
\bibitem [{\citenamefont {Takahei}\ and\ \citenamefont
  {Taguchi}(1994)}]{Takahei1994JJAP}%
  \BibitemOpen
  \bibfield  {author} {\bibinfo {author} {\bibfnamefont {K.}~\bibnamefont
  {Takahei}}\ and\ \bibinfo {author} {\bibfnamefont {A.}~\bibnamefont
  {Taguchi}},\ }\bibfield  {title} {\bibinfo {title} {{Efficient Er
  Luminescence Centers Formed in GaAs by Metalorganic Chemical Vapor Deposition
  with Oxygen Codoping}},\ }\href {https://doi.org/10.1143/JJAP.33.709}
  {\bibfield  {journal} {\bibinfo  {journal} {Jpn. J. Appl. Phys.}\ }\textbf
  {\bibinfo {volume} {33}},\ \bibinfo {pages} {709} (\bibinfo {year}
  {1994})}\BibitemShut {NoStop}%
\bibitem [{\citenamefont {Takahei}\ \emph {et~al.}(1994)\citenamefont
  {Takahei}, \citenamefont {Taguchi}, \citenamefont {Horikoshi},\ and\
  \citenamefont {Nakata}}]{Takahei1994JAP}%
  \BibitemOpen
  \bibfield  {author} {\bibinfo {author} {\bibfnamefont {K.}~\bibnamefont
  {Takahei}}, \bibinfo {author} {\bibfnamefont {A.}~\bibnamefont {Taguchi}},
  \bibinfo {author} {\bibfnamefont {Y.}~\bibnamefont {Horikoshi}},\ and\
  \bibinfo {author} {\bibfnamefont {J.}~\bibnamefont {Nakata}},\ }\bibfield
  {title} {\bibinfo {title} {{Atomic configuration of the Er‐O luminescence
  center in Er‐doped GaAs with oxygen codoping}},\ }\href
  {https://doi.org/10.1063/1.357319} {\bibfield  {journal} {\bibinfo  {journal}
  {J. Appl. Phys.}\ }\textbf {\bibinfo {volume} {76}},\ \bibinfo {pages} {4332}
  (\bibinfo {year} {1994})}\BibitemShut {NoStop}%
\bibitem [{\citenamefont {Kaczanowski}\ \emph {et~al.}(1996)\citenamefont
  {Kaczanowski}, \citenamefont {Yamamoto}, \citenamefont {Kido}, \citenamefont
  {Nakata},\ and\ \citenamefont {Takahei}}]{Kaczanowski1996NIM}%
  \BibitemOpen
  \bibfield  {author} {\bibinfo {author} {\bibfnamefont {J.}~\bibnamefont
  {Kaczanowski}}, \bibinfo {author} {\bibfnamefont {Y.}~\bibnamefont
  {Yamamoto}}, \bibinfo {author} {\bibfnamefont {Y.}~\bibnamefont {Kido}},
  \bibinfo {author} {\bibfnamefont {J.}~\bibnamefont {Nakata}},\ and\ \bibinfo
  {author} {\bibfnamefont {K.}~\bibnamefont {Takahei}},\ }\bibfield  {title}
  {\bibinfo {title} {{Monte Carlo simulation of channeling spectra for compound
  crystals with point defects and crystalline precipitates}},\ }\href
  {https://doi.org/10.1016/0168-583X(96)01460-7} {\bibfield  {journal}
  {\bibinfo  {journal} {Nucl. Instrum. Methods Phys. Res. Sect. B}\ }\textbf
  {\bibinfo {volume} {117}},\ \bibinfo {pages} {275} (\bibinfo {year}
  {1996})}\BibitemShut {NoStop}%
\bibitem [{\citenamefont {Ofuchi}\ \emph {et~al.}(2000)\citenamefont {Ofuchi},
  \citenamefont {Kubo}, \citenamefont {Tabuchi}, \citenamefont {Takahei},\ and\
  \citenamefont {Takeda}}]{Ofuchi2000ME}%
  \BibitemOpen
  \bibfield  {author} {\bibinfo {author} {\bibfnamefont {H.}~\bibnamefont
  {Ofuchi}}, \bibinfo {author} {\bibfnamefont {T.}~\bibnamefont {Kubo}},
  \bibinfo {author} {\bibfnamefont {M.}~\bibnamefont {Tabuchi}}, \bibinfo
  {author} {\bibfnamefont {K.}~\bibnamefont {Takahei}},\ and\ \bibinfo {author}
  {\bibfnamefont {Y.}~\bibnamefont {Takeda}},\ }\bibfield  {title} {\bibinfo
  {title} {Local structures around {Er} atoms in {GaAs:Er,O} studied by
  fluorescence {EXAFS} and photoluminescence},\ }\href
  {https://doi.org/10.1016/S0167-9317(99)00535-3} {\bibfield  {journal}
  {\bibinfo  {journal} {Microelectron. Eng.}\ }\textbf {\bibinfo {volume}
  {51--52}},\ \bibinfo {pages} {715} (\bibinfo {year} {2000})}\BibitemShut
  {NoStop}%
\bibitem [{\citenamefont {Takahei}\ \emph {et~al.}(1997)\citenamefont
  {Takahei}, \citenamefont {Taguchi},\ and\ \citenamefont
  {Hogg}}]{Takahei1997JAP}%
  \BibitemOpen
  \bibfield  {author} {\bibinfo {author} {\bibfnamefont {K.}~\bibnamefont
  {Takahei}}, \bibinfo {author} {\bibfnamefont {A.}~\bibnamefont {Taguchi}},\
  and\ \bibinfo {author} {\bibfnamefont {R.~A.}\ \bibnamefont {Hogg}},\
  }\bibfield  {title} {\bibinfo {title} {Atomic configurations of {Er} centers
  in {GaAs:Er,O} and {AlGaAs:Er,O} studied by site-selective luminescence
  spectroscopy},\ }\href {https://doi.org/10.1063/1.365709} {\bibfield
  {journal} {\bibinfo  {journal} {J. Appl. Phys.}\ }\textbf {\bibinfo {volume}
  {82}},\ \bibinfo {pages} {3997} (\bibinfo {year} {1997})}\BibitemShut
  {NoStop}%
\bibitem [{\citenamefont {Takahei}\ and\ \citenamefont
  {Taguchi}(1995{\natexlab{a}})}]{Takahei1995JAP}%
  \BibitemOpen
  \bibfield  {author} {\bibinfo {author} {\bibfnamefont {K.}~\bibnamefont
  {Takahei}}\ and\ \bibinfo {author} {\bibfnamefont {A.}~\bibnamefont
  {Taguchi}},\ }\bibfield  {title} {\bibinfo {title} {Photoluminescence
  analysis of {Er‐doped GaAs} under host photoexcitation and direct
  intra‐$4f$‐shell photoexcitation},\ }\href
  {https://doi.org/10.1063/1.359684} {\bibfield  {journal} {\bibinfo  {journal}
  {J. Appl. Phys.}\ }\textbf {\bibinfo {volume} {78}},\ \bibinfo {pages} {5614}
  (\bibinfo {year} {1995}{\natexlab{a}})}\BibitemShut {NoStop}%
\bibitem [{\citenamefont {Takahei}\ and\ \citenamefont
  {Taguchi}(1995{\natexlab{b}})}]{Takahei1995JAP77}%
  \BibitemOpen
  \bibfield  {author} {\bibinfo {author} {\bibfnamefont {K.}~\bibnamefont
  {Takahei}}\ and\ \bibinfo {author} {\bibfnamefont {A.}~\bibnamefont
  {Taguchi}},\ }\bibfield  {title} {\bibinfo {title}
  {Photoluminescence‐excitation analysis of {Er‐doped GaAs} grown by
  metalorganic vapor phase deposition},\ }\href
  {https://doi.org/10.1063/1.358866} {\bibfield  {journal} {\bibinfo  {journal}
  {J. Appl. Phys.}\ }\textbf {\bibinfo {volume} {77}},\ \bibinfo {pages} {1735}
  (\bibinfo {year} {1995}{\natexlab{b}})}\BibitemShut {NoStop}%
\bibitem [{\citenamefont {Ishiyama}\ \emph {et~al.}(1998)\citenamefont
  {Ishiyama}, \citenamefont {Katayama}, \citenamefont {Murakami}, \citenamefont
  {Takahei},\ and\ \citenamefont {Taguchi}}]{Ishiyama1998JAP}%
  \BibitemOpen
  \bibfield  {author} {\bibinfo {author} {\bibfnamefont {T.}~\bibnamefont
  {Ishiyama}}, \bibinfo {author} {\bibfnamefont {E.}~\bibnamefont {Katayama}},
  \bibinfo {author} {\bibfnamefont {K.}~\bibnamefont {Murakami}}, \bibinfo
  {author} {\bibfnamefont {K.}~\bibnamefont {Takahei}},\ and\ \bibinfo {author}
  {\bibfnamefont {A.}~\bibnamefont {Taguchi}},\ }\bibfield  {title} {\bibinfo
  {title} {Electron spin resonance of {Er–oxygen} complexes in {GaAs} grown
  by metal organic chemical vapor deposition},\ }\href
  {https://doi.org/10.1063/1.369009} {\bibfield  {journal} {\bibinfo  {journal}
  {J. Appl. Phys.}\ }\textbf {\bibinfo {volume} {84}},\ \bibinfo {pages} {6782}
  (\bibinfo {year} {1998})}\BibitemShut {NoStop}%
\bibitem [{\citenamefont {Haase}\ \emph {et~al.}(1996)\citenamefont {Haase},
  \citenamefont {Dornen}, \citenamefont {Takahei},\ and\ \citenamefont
  {Taguchi}}]{Haase1996MRS}%
  \BibitemOpen
  \bibfield  {author} {\bibinfo {author} {\bibfnamefont {D.}~\bibnamefont
  {Haase}}, \bibinfo {author} {\bibfnamefont {A.}~\bibnamefont {Dornen}},
  \bibinfo {author} {\bibfnamefont {K.}~\bibnamefont {Takahei}},\ and\ \bibinfo
  {author} {\bibfnamefont {A.}~\bibnamefont {Taguchi}},\ }\bibfield  {title}
  {\bibinfo {title} {{Study of the Zeeman effect of Er$^{3+}$ in GaAs:Er,O}},\
  }in\ \href@noop {} {\emph {\bibinfo {booktitle} {Rare Earth Doped
  Semiconductors II}}},\ \bibinfo {series} {Mat. Res. Soc. Symp. Proc.}, Vol.\
  \bibinfo {volume} {442}\ (\bibinfo  {publisher} {Materials Research
  Society},\ \bibinfo {address} {Pittsburgh, PA},\ \bibinfo {year} {1996})\
  pp.\ \bibinfo {pages} {179--185}\BibitemShut {NoStop}%
\bibitem [{\citenamefont {Hogg}\ \emph
  {et~al.}(1996{\natexlab{a}})\citenamefont {Hogg}, \citenamefont {Takahei},\
  and\ \citenamefont {Taguchi}}]{Hogg1996JAP}%
  \BibitemOpen
  \bibfield  {author} {\bibinfo {author} {\bibfnamefont {R.~A.}\ \bibnamefont
  {Hogg}}, \bibinfo {author} {\bibfnamefont {K.}~\bibnamefont {Takahei}},\ and\
  \bibinfo {author} {\bibfnamefont {A.}~\bibnamefont {Taguchi}},\ }\bibfield
  {title} {\bibinfo {title} {Photoluminescence excitation spectroscopy of
  {GaAs:Er,O} in the near‐band‐edge region},\ }\href
  {https://doi.org/10.1063/1.362494} {\bibfield  {journal} {\bibinfo  {journal}
  {J. Appl. Phys.}\ }\textbf {\bibinfo {volume} {79}},\ \bibinfo {pages} {8682}
  (\bibinfo {year} {1996}{\natexlab{a}})}\BibitemShut {NoStop}%
\bibitem [{\citenamefont {Hogg}\ \emph
  {et~al.}(1996{\natexlab{b}})\citenamefont {Hogg}, \citenamefont {Takahei},
  \citenamefont {Taguchi},\ and\ \citenamefont {Horikoshi}}]{Hogg1996APL}%
  \BibitemOpen
  \bibfield  {author} {\bibinfo {author} {\bibfnamefont {R.~A.}\ \bibnamefont
  {Hogg}}, \bibinfo {author} {\bibfnamefont {K.}~\bibnamefont {Takahei}},
  \bibinfo {author} {\bibfnamefont {A.}~\bibnamefont {Taguchi}},\ and\ \bibinfo
  {author} {\bibfnamefont {Y.}~\bibnamefont {Horikoshi}},\ }\bibfield  {title}
  {\bibinfo {title} {Preferential alignment of {Er–2O} centers in {GaAs:Er,O}
  revealed by anisotropic host‐excited photoluminescence},\ }\href
  {https://doi.org/10.1063/1.116043} {\bibfield  {journal} {\bibinfo  {journal}
  {Appl. Phys. Lett.}\ }\textbf {\bibinfo {volume} {68}},\ \bibinfo {pages}
  {3317} (\bibinfo {year} {1996}{\natexlab{b}})}\BibitemShut {NoStop}%
\bibitem [{\citenamefont {Culp}\ \emph {et~al.}(1998)\citenamefont {Culp},
  \citenamefont {Cederberg}, \citenamefont {Bieg}, \citenamefont {Kuech},
  \citenamefont {Bray}, \citenamefont {Pfeiffer},\ and\ \citenamefont
  {Winter}}]{Culp1998JAP}%
  \BibitemOpen
  \bibfield  {author} {\bibinfo {author} {\bibfnamefont {T.~D.}\ \bibnamefont
  {Culp}}, \bibinfo {author} {\bibfnamefont {J.~G.}\ \bibnamefont {Cederberg}},
  \bibinfo {author} {\bibfnamefont {B.}~\bibnamefont {Bieg}}, \bibinfo {author}
  {\bibfnamefont {T.~F.}\ \bibnamefont {Kuech}}, \bibinfo {author}
  {\bibfnamefont {K.~L.}\ \bibnamefont {Bray}}, \bibinfo {author}
  {\bibfnamefont {D.}~\bibnamefont {Pfeiffer}},\ and\ \bibinfo {author}
  {\bibfnamefont {C.~H.}\ \bibnamefont {Winter}},\ }\bibfield  {title}
  {\bibinfo {title} {Photoluminescence and free carrier interactions in
  erbium-doped {GaAs}},\ }\href {https://doi.org/10.1063/1.367293} {\bibfield
  {journal} {\bibinfo  {journal} {J. Appl. Phys.}\ }\textbf {\bibinfo {volume}
  {83}},\ \bibinfo {pages} {4918} (\bibinfo {year} {1998})}\BibitemShut
  {NoStop}%
\bibitem [{\citenamefont {Cederberg}\ \emph {et~al.}(1999)\citenamefont
  {Cederberg}, \citenamefont {Culp}, \citenamefont {Bieg}, \citenamefont
  {Pfeiffer}, \citenamefont {Winter}, \citenamefont {Bray},\ and\ \citenamefont
  {Kuech}}]{Cederberg1999JAP}%
  \BibitemOpen
  \bibfield  {author} {\bibinfo {author} {\bibfnamefont {J.~G.}\ \bibnamefont
  {Cederberg}}, \bibinfo {author} {\bibfnamefont {T.~D.}\ \bibnamefont {Culp}},
  \bibinfo {author} {\bibfnamefont {B.}~\bibnamefont {Bieg}}, \bibinfo {author}
  {\bibfnamefont {D.}~\bibnamefont {Pfeiffer}}, \bibinfo {author}
  {\bibfnamefont {C.~H.}\ \bibnamefont {Winter}}, \bibinfo {author}
  {\bibfnamefont {K.~L.}\ \bibnamefont {Bray}},\ and\ \bibinfo {author}
  {\bibfnamefont {T.~F.}\ \bibnamefont {Kuech}},\ }\bibfield  {title} {\bibinfo
  {title} {{Incorporation of optically active erbium into GaAs using the novel
  precursor
  tris(3,5-di-tert-butylpyrazolato)bis(4-tert-butylpyridine)erbium}},\ }\href
  {https://doi.org/10.1063/1.369179} {\bibfield  {journal} {\bibinfo  {journal}
  {J. Appl. Phys.}\ }\textbf {\bibinfo {volume} {85}},\ \bibinfo {pages} {1825}
  (\bibinfo {year} {1999})}\BibitemShut {NoStop}%
\bibitem [{\citenamefont {Fujiwara}\ \emph {et~al.}(1999)\citenamefont
  {Fujiwara}, \citenamefont {Kawamoto}, \citenamefont {Koide},\ and\
  \citenamefont {Takeda}}]{Fujiwara1999PB}%
  \BibitemOpen
  \bibfield  {author} {\bibinfo {author} {\bibfnamefont {Y.}~\bibnamefont
  {Fujiwara}}, \bibinfo {author} {\bibfnamefont {T.}~\bibnamefont {Kawamoto}},
  \bibinfo {author} {\bibfnamefont {T.}~\bibnamefont {Koide}},\ and\ \bibinfo
  {author} {\bibfnamefont {Y.}~\bibnamefont {Takeda}},\ }\bibfield  {title}
  {\bibinfo {title} {Luminescence properties of {Er,O-codoped III–V}
  semiconductors grown by organometallic vapor phase epitaxy},\ }\href
  {https://doi.org/10.1016/S0921-4526(99)00645-6} {\bibfield  {journal}
  {\bibinfo  {journal} {Phys. B: Condens. Matter}\ }\textbf {\bibinfo {volume}
  {273--274}},\ \bibinfo {pages} {770} (\bibinfo {year} {1999})}\BibitemShut
  {NoStop}%
\bibitem [{\citenamefont {Koizumi}\ \emph
  {et~al.}(2003{\natexlab{a}})\citenamefont {Koizumi}, \citenamefont
  {Fujiwara}, \citenamefont {Urakami}, \citenamefont {Inoue}, \citenamefont
  {Yoshikane},\ and\ \citenamefont {Takeda}}]{Koizumi2003PB}%
  \BibitemOpen
  \bibfield  {author} {\bibinfo {author} {\bibfnamefont {A.}~\bibnamefont
  {Koizumi}}, \bibinfo {author} {\bibfnamefont {Y.}~\bibnamefont {Fujiwara}},
  \bibinfo {author} {\bibfnamefont {A.}~\bibnamefont {Urakami}}, \bibinfo
  {author} {\bibfnamefont {K.}~\bibnamefont {Inoue}}, \bibinfo {author}
  {\bibfnamefont {T.}~\bibnamefont {Yoshikane}},\ and\ \bibinfo {author}
  {\bibfnamefont {Y.}~\bibnamefont {Takeda}},\ }\bibfield  {title} {\bibinfo
  {title} {{Effects of active layer thickness on Er excitation cross section in
  GaInP/GaAs:Er,O/GaInP double heterostructure light-emitting diodes}},\ }\href
  {https://doi.org/10.1016/j.physb.2003.09.085} {\bibfield  {journal} {\bibinfo
   {journal} {Phys. B: Condens. Matter}\ }\textbf {\bibinfo {volume}
  {340-342}},\ \bibinfo {pages} {309} (\bibinfo {year}
  {2003}{\natexlab{a}})}\BibitemShut {NoStop}%
\bibitem [{\citenamefont {Koizumi}\ \emph
  {et~al.}(2003{\natexlab{b}})\citenamefont {Koizumi}, \citenamefont
  {Fujiwara}, \citenamefont {Urakami}, \citenamefont {Inoue}, \citenamefont
  {Yoshikane},\ and\ \citenamefont {Takeda}}]{Koizumi2003APL}%
  \BibitemOpen
  \bibfield  {author} {\bibinfo {author} {\bibfnamefont {A.}~\bibnamefont
  {Koizumi}}, \bibinfo {author} {\bibfnamefont {Y.}~\bibnamefont {Fujiwara}},
  \bibinfo {author} {\bibfnamefont {A.}~\bibnamefont {Urakami}}, \bibinfo
  {author} {\bibfnamefont {K.}~\bibnamefont {Inoue}}, \bibinfo {author}
  {\bibfnamefont {T.}~\bibnamefont {Yoshikane}},\ and\ \bibinfo {author}
  {\bibfnamefont {Y.}~\bibnamefont {Takeda}},\ }\bibfield  {title} {\bibinfo
  {title} {Room-temperature electroluminescence properties of {Er,O-codoped
  GaAs} injection-type light-emitting diodes grown by organometallic vapor
  phase epitaxy},\ }\href {https://doi.org/10.1063/1.1630165} {\bibfield
  {journal} {\bibinfo  {journal} {Appl. Phys. Lett.}\ }\textbf {\bibinfo
  {volume} {83}},\ \bibinfo {pages} {4521} (\bibinfo {year}
  {2003}{\natexlab{b}})}\BibitemShut {NoStop}%
\bibitem [{\citenamefont {Fujiwara}(2005)}]{Fujiwara2005MT}%
  \BibitemOpen
  \bibfield  {author} {\bibinfo {author} {\bibfnamefont {Y.}~\bibnamefont
  {Fujiwara}},\ }\bibfield  {title} {\bibinfo {title} {{Room-Temperature
  Operation of Injection-Type 1.5 $\mu$m Light-Emitting Diodes with
  Er,O-Codoped GaAs}},\ }\href {https://doi.org/10.2320/matertrans.46.1969}
  {\bibfield  {journal} {\bibinfo  {journal} {Mater. Trans.}\ }\textbf
  {\bibinfo {volume} {46}},\ \bibinfo {pages} {1969} (\bibinfo {year}
  {2005})}\BibitemShut {NoStop}%
\bibitem [{\citenamefont {Taguchi}\ and\ \citenamefont
  {Takahei}(1996)}]{Taguchi1996JAP}%
  \BibitemOpen
  \bibfield  {author} {\bibinfo {author} {\bibfnamefont {A.}~\bibnamefont
  {Taguchi}}\ and\ \bibinfo {author} {\bibfnamefont {K.}~\bibnamefont
  {Takahei}},\ }\bibfield  {title} {\bibinfo {title} {{Trap level
  characteristics of rare‐earth luminescence centers in III-V
  semiconductors}},\ }\href {https://doi.org/10.1063/1.361741} {\bibfield
  {journal} {\bibinfo  {journal} {J. Appl. Phys.}\ }\textbf {\bibinfo {volume}
  {79}},\ \bibinfo {pages} {4330} (\bibinfo {year} {1996})}\BibitemShut
  {NoStop}%
\bibitem [{\citenamefont {Takahei}\ \emph {et~al.}(1996)\citenamefont
  {Takahei}, \citenamefont {Hogg},\ and\ \citenamefont
  {Taguchi}}]{Takahei1996MRS}%
  \BibitemOpen
  \bibfield  {author} {\bibinfo {author} {\bibfnamefont {K.}~\bibnamefont
  {Takahei}}, \bibinfo {author} {\bibfnamefont {R.}~\bibnamefont {Hogg}},\ and\
  \bibinfo {author} {\bibfnamefont {A.}~\bibnamefont {Taguchi}},\ }\bibfield
  {title} {\bibinfo {title} {Energy-transfer processes in oxygen-coped
  {GaAs:Er}},\ }in\ \href@noop {} {\emph {\bibinfo {booktitle} {Rare Earth
  Doped Semiconductors II}}},\ \bibinfo {series} {Mat. Res. Soc. Symp. Proc.},
  Vol.\ \bibinfo {volume} {442}\ (\bibinfo  {publisher} {Materials Research
  Society},\ \bibinfo {address} {Pittsburgh, PA},\ \bibinfo {year} {1996})\
  pp.\ \bibinfo {pages} {267--278}\BibitemShut {NoStop}%
\bibitem [{\citenamefont {Hogg}\ \emph {et~al.}(1997)\citenamefont {Hogg},
  \citenamefont {Takahei},\ and\ \citenamefont {Taguchi}}]{Hogg1997PRB}%
  \BibitemOpen
  \bibfield  {author} {\bibinfo {author} {\bibfnamefont {R.~A.}\ \bibnamefont
  {Hogg}}, \bibinfo {author} {\bibfnamefont {K.}~\bibnamefont {Takahei}},\ and\
  \bibinfo {author} {\bibfnamefont {A.}~\bibnamefont {Taguchi}},\ }\bibfield
  {title} {\bibinfo {title} {Er-related trap levels in {GaAs:Er,O} studied by
  optical spectroscopy under hydrostatic pressure},\ }\href
  {https://doi.org/10.1103/PhysRevB.56.10255} {\bibfield  {journal} {\bibinfo
  {journal} {Phys. Rev. B}\ }\textbf {\bibinfo {volume} {56}},\ \bibinfo
  {pages} {10255} (\bibinfo {year} {1997})}\BibitemShut {NoStop}%
\bibitem [{\citenamefont {Nakamura}\ \emph {et~al.}(2006)\citenamefont
  {Nakamura}, \citenamefont {Takemoto}, \citenamefont {Terai}, \citenamefont
  {Suzuki}, \citenamefont {Koizumi}, \citenamefont {Takeda}, \citenamefont
  {Tonouchi},\ and\ \citenamefont {Fujiwara}}]{Nakamura2006PB}%
  \BibitemOpen
  \bibfield  {author} {\bibinfo {author} {\bibfnamefont {K.}~\bibnamefont
  {Nakamura}}, \bibinfo {author} {\bibfnamefont {S.}~\bibnamefont {Takemoto}},
  \bibinfo {author} {\bibfnamefont {Y.}~\bibnamefont {Terai}}, \bibinfo
  {author} {\bibfnamefont {M.}~\bibnamefont {Suzuki}}, \bibinfo {author}
  {\bibfnamefont {A.}~\bibnamefont {Koizumi}}, \bibinfo {author} {\bibfnamefont
  {Y.}~\bibnamefont {Takeda}}, \bibinfo {author} {\bibfnamefont
  {M.}~\bibnamefont {Tonouchi}},\ and\ \bibinfo {author} {\bibfnamefont
  {Y.}~\bibnamefont {Fujiwara}},\ }\bibfield  {title} {\bibinfo {title} {Direct
  observation of trapping of photo-excited carriers in {Er,O-codoped GaAs}},\
  }\href {https://doi.org/10.1016/j.physb.2005.12.140} {\bibfield  {journal}
  {\bibinfo  {journal} {Phys. B: Condens. Matter}\ }\textbf {\bibinfo {volume}
  {376--377}},\ \bibinfo {pages} {556} (\bibinfo {year} {2006})}\BibitemShut
  {NoStop}%
\bibitem [{\citenamefont {Ishii}\ \emph {et~al.}(2014)\citenamefont {Ishii},
  \citenamefont {Koizumi}, \citenamefont {Takeda},\ and\ \citenamefont
  {Fujiwara}}]{Ishii2014JAP}%
  \BibitemOpen
  \bibfield  {author} {\bibinfo {author} {\bibfnamefont {M.}~\bibnamefont
  {Ishii}}, \bibinfo {author} {\bibfnamefont {A.}~\bibnamefont {Koizumi}},
  \bibinfo {author} {\bibfnamefont {Y.}~\bibnamefont {Takeda}},\ and\ \bibinfo
  {author} {\bibfnamefont {Y.}~\bibnamefont {Fujiwara}},\ }\bibfield  {title}
  {\bibinfo {title} {Discrimination between energy transfer and back transfer
  processes for {GaAs} host and {Er} luminescent dopants using electric
  response analysis},\ }\href {https://doi.org/10.1063/1.4870808} {\bibfield
  {journal} {\bibinfo  {journal} {J. Appl. Phys.}\ }\textbf {\bibinfo {volume}
  {115}},\ \bibinfo {pages} {133510} (\bibinfo {year} {2014})}\BibitemShut
  {NoStop}%
\bibitem [{\citenamefont {Katsuno}\ \emph {et~al.}(2011)\citenamefont
  {Katsuno}, \citenamefont {Ohta}, \citenamefont {Portugall}, \citenamefont
  {Ubrig}, \citenamefont {Fujisawa}, \citenamefont {Elmasry}, \citenamefont
  {Okubo},\ and\ \citenamefont {Fujiwara}}]{Katsuno2011JL}%
  \BibitemOpen
  \bibfield  {author} {\bibinfo {author} {\bibfnamefont {H.}~\bibnamefont
  {Katsuno}}, \bibinfo {author} {\bibfnamefont {H.}~\bibnamefont {Ohta}},
  \bibinfo {author} {\bibfnamefont {O.}~\bibnamefont {Portugall}}, \bibinfo
  {author} {\bibfnamefont {N.}~\bibnamefont {Ubrig}}, \bibinfo {author}
  {\bibfnamefont {M.}~\bibnamefont {Fujisawa}}, \bibinfo {author}
  {\bibfnamefont {F.}~\bibnamefont {Elmasry}}, \bibinfo {author} {\bibfnamefont
  {S.}~\bibnamefont {Okubo}},\ and\ \bibinfo {author} {\bibfnamefont
  {Y.}~\bibnamefont {Fujiwara}},\ }\bibfield  {title} {\bibinfo {title} {Energy
  structure of {Er-2O} center in {GaAs:Er,O} studied by high magnetic field
  photoluminescence measurement},\ }\href
  {https://doi.org/10.1016/j.jlumin.2011.05.034} {\bibfield  {journal}
  {\bibinfo  {journal} {J. Lumin.}\ }\textbf {\bibinfo {volume} {131}},\
  \bibinfo {pages} {2294} (\bibinfo {year} {2011})}\BibitemShut {NoStop}%
\bibitem [{\citenamefont {Elmasry}\ \emph {et~al.}(2014)\citenamefont
  {Elmasry}, \citenamefont {Okubo}, \citenamefont {Ohta},\ and\ \citenamefont
  {Fujiwara}}]{Elmasry2014JAP}%
  \BibitemOpen
  \bibfield  {author} {\bibinfo {author} {\bibfnamefont {F.}~\bibnamefont
  {Elmasry}}, \bibinfo {author} {\bibfnamefont {S.}~\bibnamefont {Okubo}},
  \bibinfo {author} {\bibfnamefont {H.}~\bibnamefont {Ohta}},\ and\ \bibinfo
  {author} {\bibfnamefont {Y.}~\bibnamefont {Fujiwara}},\ }\bibfield  {title}
  {\bibinfo {title} {{Electron spin resonance study of Er-concentration effect
  in GaAs:Er,O containing charge carriers}},\ }\href
  {https://doi.org/10.1063/1.4876487} {\bibfield  {journal} {\bibinfo
  {journal} {J. Appl. Phys.}\ }\textbf {\bibinfo {volume} {115}},\ \bibinfo
  {pages} {193904} (\bibinfo {year} {2014})}\BibitemShut {NoStop}%
\bibitem [{\citenamefont {Maltez}\ \emph {et~al.}(2004)\citenamefont {Maltez},
  \citenamefont {Ribeiro}, \citenamefont {Bernussi}, \citenamefont {Amaral},
  \citenamefont {Behar}, \citenamefont {Specht},\ and\ \citenamefont
  {Liliental-Weber}}]{Maltez2004NIMPR}%
  \BibitemOpen
  \bibfield  {author} {\bibinfo {author} {\bibfnamefont {R.~L.}\ \bibnamefont
  {Maltez}}, \bibinfo {author} {\bibfnamefont {E.}~\bibnamefont {Ribeiro}},
  \bibinfo {author} {\bibfnamefont {A.~A.}\ \bibnamefont {Bernussi}}, \bibinfo
  {author} {\bibfnamefont {L.}~\bibnamefont {Amaral}}, \bibinfo {author}
  {\bibfnamefont {M.}~\bibnamefont {Behar}}, \bibinfo {author} {\bibfnamefont
  {P.}~\bibnamefont {Specht}},\ and\ \bibinfo {author} {\bibfnamefont
  {Z.}~\bibnamefont {Liliental-Weber}},\ }\bibfield  {title} {\bibinfo {title}
  {{TEM and PL characterization of erbium and oxygen co-implanted
  LT-GaAs:Be}},\ }\href {https://doi.org/10.1016/j.nimb.2003.12.061} {\bibfield
   {journal} {\bibinfo  {journal} {Nucl. Instrum. Methods Phys. Res., Sect. B}\
  }\textbf {\bibinfo {volume} {218}},\ \bibinfo {pages} {444} (\bibinfo {year}
  {2004})}\BibitemShut {NoStop}%
\bibitem [{\citenamefont {Gregorkiewicz}\ and\ \citenamefont
  {Langer}(1999)}]{Gregorkiewicz1999MRSBulletin}%
  \BibitemOpen
  \bibfield  {author} {\bibinfo {author} {\bibfnamefont {T.}~\bibnamefont
  {Gregorkiewicz}}\ and\ \bibinfo {author} {\bibfnamefont {J.}~\bibnamefont
  {Langer}},\ }\bibfield  {title} {\bibinfo {title} {{Lasing in
  Rare-Earth-Doped Semiconductors: Hopes and Facts}},\ }\href
  {https://doi.org/10.1557/S0883769400053033} {\bibfield  {journal} {\bibinfo
  {journal} {MRS Bulletin}\ }\textbf {\bibinfo {volume} {24}},\ \bibinfo
  {pages} {27–32} (\bibinfo {year} {1999})}\BibitemShut {NoStop}%
\bibitem [{\citenamefont {Hoang}(2015)}]{Hoang2015RRL}%
  \BibitemOpen
  \bibfield  {author} {\bibinfo {author} {\bibfnamefont {K.}~\bibnamefont
  {Hoang}},\ }\bibfield  {title} {\bibinfo {title} {{Hybrid density functional
  study of optically active Er$^{3+}$ centers in GaN}},\ }\href
  {https://doi.org/10.1002/pssr.201510269} {\bibfield  {journal} {\bibinfo
  {journal} {Phys. Status Solidi RRL}\ }\textbf {\bibinfo {volume} {9}},\
  \bibinfo {pages} {722} (\bibinfo {year} {2015})}\BibitemShut {NoStop}%
\bibitem [{\citenamefont {Hoang}(2016)}]{Hoang2016RRL}%
  \BibitemOpen
  \bibfield  {author} {\bibinfo {author} {\bibfnamefont {K.}~\bibnamefont
  {Hoang}},\ }\bibfield  {title} {\bibinfo {title} {{First-principles
  identification of defect levels in Er-doped GaN}},\ }\href
  {https://doi.org/10.1002/pssr.201600273} {\bibfield  {journal} {\bibinfo
  {journal} {Phys. Status Solidi RRL}\ }\textbf {\bibinfo {volume} {10}},\
  \bibinfo {pages} {915} (\bibinfo {year} {2016})}\BibitemShut {NoStop}%
\bibitem [{\citenamefont {Hoang}(2022{\natexlab{a}})}]{Hoang201JAP}%
  \BibitemOpen
  \bibfield  {author} {\bibinfo {author} {\bibfnamefont {K.}~\bibnamefont
  {Hoang}},\ }\bibfield  {title} {\bibinfo {title} {Rare-earth defects and
  defect-related luminescence in {ZnS}},\ }\href
  {https://doi.org/10.1063/5.0069390} {\bibfield  {journal} {\bibinfo
  {journal} {J. Appl. Phys.}\ }\textbf {\bibinfo {volume} {131}},\ \bibinfo
  {pages} {015705} (\bibinfo {year} {2022}{\natexlab{a}})}\BibitemShut
  {NoStop}%
\bibitem [{\citenamefont {Hoang}(2021)}]{Hoang2021PRM}%
  \BibitemOpen
  \bibfield  {author} {\bibinfo {author} {\bibfnamefont {K.}~\bibnamefont
  {Hoang}},\ }\bibfield  {title} {\bibinfo {title} {Tuning the valence and
  concentration of europium and luminescence centers in {GaN} through co-doping
  and defect association},\ }\href
  {https://doi.org/10.1103/PhysRevMaterials.5.034601} {\bibfield  {journal}
  {\bibinfo  {journal} {Phys. Rev. Materials}\ }\textbf {\bibinfo {volume}
  {5}},\ \bibinfo {pages} {034601} (\bibinfo {year} {2021})}\BibitemShut
  {NoStop}%
\bibitem [{\citenamefont {Hoang}(2022{\natexlab{b}})}]{Hoang2022PRM}%
  \BibitemOpen
  \bibfield  {author} {\bibinfo {author} {\bibfnamefont {K.}~\bibnamefont
  {Hoang}},\ }\bibfield  {title} {\bibinfo {title} {Rare-earth defects in
  {GaN}: {A} systematic investigation of the lanthanide series},\ }\href
  {https://doi.org/10.1103/PhysRevMaterials.6.044601} {\bibfield  {journal}
  {\bibinfo  {journal} {Phys. Rev. Materials}\ }\textbf {\bibinfo {volume}
  {6}},\ \bibinfo {pages} {044601} (\bibinfo {year}
  {2022}{\natexlab{b}})}\BibitemShut {NoStop}%
\bibitem [{\citenamefont {Taguchi}\ and\ \citenamefont
  {Ohno}(1997)}]{Taguchi1997PRB}%
  \BibitemOpen
  \bibfield  {author} {\bibinfo {author} {\bibfnamefont {A.}~\bibnamefont
  {Taguchi}}\ and\ \bibinfo {author} {\bibfnamefont {T.}~\bibnamefont {Ohno}},\
  }\bibfield  {title} {\bibinfo {title} {{Erbium in GaAs: Coupling with native
  defects}},\ }\href {https://doi.org/10.1103/PhysRevB.56.9477} {\bibfield
  {journal} {\bibinfo  {journal} {Phys. Rev. B}\ }\textbf {\bibinfo {volume}
  {56}},\ \bibinfo {pages} {9477} (\bibinfo {year} {1997})}\BibitemShut
  {NoStop}%
\bibitem [{\citenamefont {Svane}\ \emph {et~al.}(2006)\citenamefont {Svane},
  \citenamefont {Christensen}, \citenamefont {Petit}, \citenamefont {Szotek},\
  and\ \citenamefont {Temmerman}}]{Svane2006}%
  \BibitemOpen
  \bibfield  {author} {\bibinfo {author} {\bibfnamefont {A.}~\bibnamefont
  {Svane}}, \bibinfo {author} {\bibfnamefont {N.~E.}\ \bibnamefont
  {Christensen}}, \bibinfo {author} {\bibfnamefont {L.}~\bibnamefont {Petit}},
  \bibinfo {author} {\bibfnamefont {Z.}~\bibnamefont {Szotek}},\ and\ \bibinfo
  {author} {\bibfnamefont {W.~M.}\ \bibnamefont {Temmerman}},\ }\bibfield
  {title} {\bibinfo {title} {{Electronic structure of rare-earth impurities in
  GaAs and GaN}},\ }\href {https://doi.org/10.1103/PhysRevB.74.165204}
  {\bibfield  {journal} {\bibinfo  {journal} {Phys. Rev. B}\ }\textbf {\bibinfo
  {volume} {74}},\ \bibinfo {pages} {165204} (\bibinfo {year}
  {2006})}\BibitemShut {NoStop}%
\bibitem [{\citenamefont {Coutinho}\ \emph {et~al.}(2004)\citenamefont
  {Coutinho}, \citenamefont {Jones}, \citenamefont {Shaw}, \citenamefont
  {Briddon},\ and\ \citenamefont {Öberg}}]{Coutinho2004APL}%
  \BibitemOpen
  \bibfield  {author} {\bibinfo {author} {\bibfnamefont {J.}~\bibnamefont
  {Coutinho}}, \bibinfo {author} {\bibfnamefont {R.}~\bibnamefont {Jones}},
  \bibinfo {author} {\bibfnamefont {M.~J.}\ \bibnamefont {Shaw}}, \bibinfo
  {author} {\bibfnamefont {P.~R.}\ \bibnamefont {Briddon}},\ and\ \bibinfo
  {author} {\bibfnamefont {S.}~\bibnamefont {Öberg}},\ }\bibfield  {title}
  {\bibinfo {title} {Optically active erbium–oxygen complexes in {GaAs}},\
  }\href {https://doi.org/10.1063/1.1668323} {\bibfield  {journal} {\bibinfo
  {journal} {Appl. Phys. Lett.}\ }\textbf {\bibinfo {volume} {84}},\ \bibinfo
  {pages} {1683} (\bibinfo {year} {2004})}\BibitemShut {NoStop}%
\bibitem [{\citenamefont {Freysoldt}\ \emph {et~al.}(2014)\citenamefont
  {Freysoldt}, \citenamefont {Grabowski}, \citenamefont {Hickel}, \citenamefont
  {Neugebauer}, \citenamefont {Kresse}, \citenamefont {Janotti},\ and\
  \citenamefont {{Van de Walle}}}]{Freysoldt2014RMP}%
  \BibitemOpen
  \bibfield  {author} {\bibinfo {author} {\bibfnamefont {C.}~\bibnamefont
  {Freysoldt}}, \bibinfo {author} {\bibfnamefont {B.}~\bibnamefont
  {Grabowski}}, \bibinfo {author} {\bibfnamefont {T.}~\bibnamefont {Hickel}},
  \bibinfo {author} {\bibfnamefont {J.}~\bibnamefont {Neugebauer}}, \bibinfo
  {author} {\bibfnamefont {G.}~\bibnamefont {Kresse}}, \bibinfo {author}
  {\bibfnamefont {A.}~\bibnamefont {Janotti}},\ and\ \bibinfo {author}
  {\bibfnamefont {C.~G.}\ \bibnamefont {{Van de Walle}}},\ }\bibfield  {title}
  {\bibinfo {title} {First-principles calculations for point defects in
  solids},\ }\href {https://doi.org/10.1103/RevModPhys.86.253} {\bibfield
  {journal} {\bibinfo  {journal} {Rev. Mod. Phys.}\ }\textbf {\bibinfo {volume}
  {86}},\ \bibinfo {pages} {253} (\bibinfo {year} {2014})}\BibitemShut
  {NoStop}%
\bibitem [{\citenamefont {Hoang}(2026)}]{Hoang2026JPCM}%
  \BibitemOpen
  \bibfield  {author} {\bibinfo {author} {\bibfnamefont {K.}~\bibnamefont
  {Hoang}},\ }\bibfield  {title} {\bibinfo {title} {First-principles
  characterization of native defects and oxygen impurities in {GaAs}},\ }\href
  {https://doi.org/10.1088/1361-648X/ae7245} {\bibfield  {journal} {\bibinfo
  {journal} {J. Phys.: Condens. Matter}\ }\textbf {\bibinfo {volume} {38}},\
  \bibinfo {pages} {225701} (\bibinfo {year} {2026})}\BibitemShut {NoStop}%
\bibitem [{\citenamefont {Alkauskas}\ \emph {et~al.}(2014)\citenamefont
  {Alkauskas}, \citenamefont {Yan},\ and\ \citenamefont {Van~de
  Walle}}]{Alkauskas2014PRB}%
  \BibitemOpen
  \bibfield  {author} {\bibinfo {author} {\bibfnamefont {A.}~\bibnamefont
  {Alkauskas}}, \bibinfo {author} {\bibfnamefont {Q.}~\bibnamefont {Yan}},\
  and\ \bibinfo {author} {\bibfnamefont {C.~G.}\ \bibnamefont {Van~de Walle}},\
  }\bibfield  {title} {\bibinfo {title} {First-principles theory of
  nonradiative carrier capture via multiphonon emission},\ }\href
  {https://doi.org/10.1103/PhysRevB.90.075202} {\bibfield  {journal} {\bibinfo
  {journal} {Phys. Rev. B}\ }\textbf {\bibinfo {volume} {90}},\ \bibinfo
  {pages} {075202} (\bibinfo {year} {2014})}\BibitemShut {NoStop}%
\bibitem [{\citenamefont {Turiansky}\ \emph {et~al.}(2021)\citenamefont
  {Turiansky}, \citenamefont {Alkauskas}, \citenamefont {Engel}, \citenamefont
  {Kresse}, \citenamefont {Wickramaratne}, \citenamefont {Shen}, \citenamefont
  {Dreyer},\ and\ \citenamefont {{Van de Walle}}}]{Turiansky2021CPC}%
  \BibitemOpen
  \bibfield  {author} {\bibinfo {author} {\bibfnamefont {M.~E.}\ \bibnamefont
  {Turiansky}}, \bibinfo {author} {\bibfnamefont {A.}~\bibnamefont
  {Alkauskas}}, \bibinfo {author} {\bibfnamefont {M.}~\bibnamefont {Engel}},
  \bibinfo {author} {\bibfnamefont {G.}~\bibnamefont {Kresse}}, \bibinfo
  {author} {\bibfnamefont {D.}~\bibnamefont {Wickramaratne}}, \bibinfo {author}
  {\bibfnamefont {J.-X.}\ \bibnamefont {Shen}}, \bibinfo {author}
  {\bibfnamefont {C.~E.}\ \bibnamefont {Dreyer}},\ and\ \bibinfo {author}
  {\bibfnamefont {C.~G.}\ \bibnamefont {{Van de Walle}}},\ }\bibfield  {title}
  {\bibinfo {title} {Nonrad: {Computing} nonradiative capture coefficients from
  first principles},\ }\href {https://doi.org/10.1016/j.cpc.2021.108056}
  {\bibfield  {journal} {\bibinfo  {journal} {Comput. Phys. Commun.}\ }\textbf
  {\bibinfo {volume} {267}},\ \bibinfo {pages} {108056} (\bibinfo {year}
  {2021})}\BibitemShut {NoStop}%
\bibitem [{\citenamefont {Heyd}\ \emph {et~al.}(2003)\citenamefont {Heyd},
  \citenamefont {Scuseria},\ and\ \citenamefont {Ernzerhof}}]{heyd:8207}%
  \BibitemOpen
  \bibfield  {author} {\bibinfo {author} {\bibfnamefont {J.}~\bibnamefont
  {Heyd}}, \bibinfo {author} {\bibfnamefont {G.~E.}\ \bibnamefont {Scuseria}},\
  and\ \bibinfo {author} {\bibfnamefont {M.}~\bibnamefont {Ernzerhof}},\
  }\bibfield  {title} {\bibinfo {title} {{Hybrid functionals based on a
  screened Coulomb potential}},\ }\href {https://doi.org/10.1063/1.1564060}
  {\bibfield  {journal} {\bibinfo  {journal} {J. Chem. Phys.}\ }\textbf
  {\bibinfo {volume} {118}},\ \bibinfo {pages} {8207} (\bibinfo {year}
  {2003})}\BibitemShut {NoStop}%
\bibitem [{\citenamefont {Kresse}\ and\ \citenamefont {Joubert}(1999)}]{PAW2}%
  \BibitemOpen
  \bibfield  {author} {\bibinfo {author} {\bibfnamefont {G.}~\bibnamefont
  {Kresse}}\ and\ \bibinfo {author} {\bibfnamefont {D.}~\bibnamefont
  {Joubert}},\ }\bibfield  {title} {\bibinfo {title} {From ultrasoft
  pseudopotentials to the projector augmented-wave method},\ }\href
  {https://doi.org/10.1103/PhysRevB.59.1758} {\bibfield  {journal} {\bibinfo
  {journal} {Phys. Rev. B}\ }\textbf {\bibinfo {volume} {59}},\ \bibinfo
  {pages} {1758} (\bibinfo {year} {1999})}\BibitemShut {NoStop}%
\bibitem [{\citenamefont {Kresse}\ and\ \citenamefont
  {Furthm{\"u}ller}(1996)}]{VASP2}%
  \BibitemOpen
  \bibfield  {author} {\bibinfo {author} {\bibfnamefont {G.}~\bibnamefont
  {Kresse}}\ and\ \bibinfo {author} {\bibfnamefont {J.}~\bibnamefont
  {Furthm{\"u}ller}},\ }\bibfield  {title} {\bibinfo {title} {Efficient
  iterative schemes for ab initio total-energy calculations using a plane-wave
  basis set},\ }\href {https://doi.org/10.1103/PhysRevB.54.11169} {\bibfield
  {journal} {\bibinfo  {journal} {Phys. Rev. B}\ }\textbf {\bibinfo {volume}
  {54}},\ \bibinfo {pages} {11169} (\bibinfo {year} {1996})}\BibitemShut
  {NoStop}%
\bibitem [{\citenamefont {{Van de Walle}}\ and\ \citenamefont
  {Neugebauer}(2004)}]{walle:3851}%
  \BibitemOpen
  \bibfield  {author} {\bibinfo {author} {\bibfnamefont {C.~G.}\ \bibnamefont
  {{Van de Walle}}}\ and\ \bibinfo {author} {\bibfnamefont {J.}~\bibnamefont
  {Neugebauer}},\ }\bibfield  {title} {\bibinfo {title} {{First-principles
  calculations for defects and impurities: Applications to III-nitrides}},\
  }\href {https://doi.org/10.1063/1.1682673} {\bibfield  {journal} {\bibinfo
  {journal} {J. Appl. Phys.}\ }\textbf {\bibinfo {volume} {95}},\ \bibinfo
  {pages} {3851} (\bibinfo {year} {2004})}\BibitemShut {NoStop}%
\bibitem [{\citenamefont {Freysoldt}\ \emph {et~al.}(2009)\citenamefont
  {Freysoldt}, \citenamefont {Neugebauer},\ and\ \citenamefont {{Van de
  Walle}}}]{Freysoldt}%
  \BibitemOpen
  \bibfield  {author} {\bibinfo {author} {\bibfnamefont {C.}~\bibnamefont
  {Freysoldt}}, \bibinfo {author} {\bibfnamefont {J.}~\bibnamefont
  {Neugebauer}},\ and\ \bibinfo {author} {\bibfnamefont {C.~G.}\ \bibnamefont
  {{Van de Walle}}},\ }\bibfield  {title} {\bibinfo {title} {{Fully \textit{Ab
  Initio} Finite-Size Corrections for Charged-Defect Supercell Calculations}},\
  }\href {https://doi.org/10.1103/PhysRevLett.102.016402} {\bibfield  {journal}
  {\bibinfo  {journal} {Phys. Rev. Lett.}\ }\textbf {\bibinfo {volume} {102}},\
  \bibinfo {pages} {016402} (\bibinfo {year} {2009})}\BibitemShut {NoStop}%
\bibitem [{\citenamefont {Freysoldt}\ \emph {et~al.}(2011)\citenamefont
  {Freysoldt}, \citenamefont {Neugebauer},\ and\ \citenamefont {{Van de
  Walle}}}]{Freysoldt11}%
  \BibitemOpen
  \bibfield  {author} {\bibinfo {author} {\bibfnamefont {C.}~\bibnamefont
  {Freysoldt}}, \bibinfo {author} {\bibfnamefont {J.}~\bibnamefont
  {Neugebauer}},\ and\ \bibinfo {author} {\bibfnamefont {C.~G.}\ \bibnamefont
  {{Van de Walle}}},\ }\bibfield  {title} {\bibinfo {title} {Electrostatic
  interactions between charged defects in supercells},\ }\href
  {https://doi.org/10.1002/pssb.201046289} {\bibfield  {journal} {\bibinfo
  {journal} {phys. status solidi (b)}\ }\textbf {\bibinfo {volume} {248}},\
  \bibinfo {pages} {1067} (\bibinfo {year} {2011})}\BibitemShut {NoStop}%
\bibitem [{\citenamefont {Hanks}\ and\ \citenamefont
  {Faktor}(1967)}]{Hanks1967TFS}%
  \BibitemOpen
  \bibfield  {author} {\bibinfo {author} {\bibfnamefont {R.}~\bibnamefont
  {Hanks}}\ and\ \bibinfo {author} {\bibfnamefont {M.~M.}\ \bibnamefont
  {Faktor}},\ }\bibfield  {title} {\bibinfo {title} {{Quantitative application
  of dynamic differential calorimetry. Part 2.--Heats of formation of the group
  3A arsenides}},\ }\href {https://doi.org/10.1039/TF9676301130} {\bibfield
  {journal} {\bibinfo  {journal} {Trans. Faraday Soc.}\ }\textbf {\bibinfo
  {volume} {63}},\ \bibinfo {pages} {1130} (\bibinfo {year}
  {1967})}\BibitemShut {NoStop}%
\bibitem [{\citenamefont {Monkhorst}\ and\ \citenamefont
  {Pack}(1976)}]{monkhorst-pack}%
  \BibitemOpen
  \bibfield  {author} {\bibinfo {author} {\bibfnamefont {H.~J.}\ \bibnamefont
  {Monkhorst}}\ and\ \bibinfo {author} {\bibfnamefont {J.~D.}\ \bibnamefont
  {Pack}},\ }\bibfield  {title} {\bibinfo {title} {{Special points for
  Brillouin-zone integrations}},\ }\href
  {https://doi.org/10.1103/PhysRevB.13.5188} {\bibfield  {journal} {\bibinfo
  {journal} {Phys. Rev. B}\ }\textbf {\bibinfo {volume} {13}},\ \bibinfo
  {pages} {5188} (\bibinfo {year} {1976})}\BibitemShut {NoStop}%
\bibitem [{\citenamefont {Wang}\ \emph {et~al.}(2021)\citenamefont {Wang},
  \citenamefont {Xu}, \citenamefont {Liu}, \citenamefont {Tang},\ and\
  \citenamefont {Geng}}]{Wang2021CPC}%
  \BibitemOpen
  \bibfield  {author} {\bibinfo {author} {\bibfnamefont {V.}~\bibnamefont
  {Wang}}, \bibinfo {author} {\bibfnamefont {N.}~\bibnamefont {Xu}}, \bibinfo
  {author} {\bibfnamefont {J.-C.}\ \bibnamefont {Liu}}, \bibinfo {author}
  {\bibfnamefont {G.}~\bibnamefont {Tang}},\ and\ \bibinfo {author}
  {\bibfnamefont {W.-T.}\ \bibnamefont {Geng}},\ }\bibfield  {title} {\bibinfo
  {title} {{VASPKIT}: {A} user-friendly interface facilitating high-throughput
  computing and analysis using {VASP} code},\ }\href
  {https://doi.org/10.1016/j.cpc.2021.108033} {\bibfield  {journal} {\bibinfo
  {journal} {Comput. Phys. Commun.}\ }\textbf {\bibinfo {volume} {267}},\
  \bibinfo {pages} {108033} (\bibinfo {year} {2021})}\BibitemShut {NoStop}%
\bibitem [{\citenamefont {Shannon}(1976)}]{Shannon1976}%
  \BibitemOpen
  \bibfield  {author} {\bibinfo {author} {\bibfnamefont {R.~D.}\ \bibnamefont
  {Shannon}},\ }\bibfield  {title} {\bibinfo {title} {{Revised effective ionic
  radii and systematic studies of interatomic distances in halides and
  chalcogenides}},\ }\href {https://doi.org/10.1107/S0567739476001551}
  {\bibfield  {journal} {\bibinfo  {journal} {Acta Crystallogr., Sect. A:
  Found. Crystallogr.}\ }\textbf {\bibinfo {volume} {32}},\ \bibinfo {pages}
  {751} (\bibinfo {year} {1976})}\BibitemShut {NoStop}%
\bibitem [{\citenamefont {Fiorentini}(1995)}]{Fiorentini1995PRB}%
  \BibitemOpen
  \bibfield  {author} {\bibinfo {author} {\bibfnamefont {V.}~\bibnamefont
  {Fiorentini}},\ }\bibfield  {title} {\bibinfo {title} {Effective-mass single
  and double acceptor spectra in {GaAs}},\ }\href
  {https://doi.org/10.1103/PhysRevB.51.10161} {\bibfield  {journal} {\bibinfo
  {journal} {Phys. Rev. B}\ }\textbf {\bibinfo {volume} {51}},\ \bibinfo
  {pages} {10161} (\bibinfo {year} {1995})}\BibitemShut {NoStop}%
\end{thebibliography}

%

\end{document}